\documentclass{emulateapj}

    \usepackage{graphicx}
    \usepackage{adjustbox}
    \usepackage{enumerate} 
    \usepackage{amsmath} 
    \usepackage{amssymb} 
    \usepackage[mathletters]{ucs} 
    \usepackage[utf8x]{inputenc} 
    \usepackage{hyperref}
    \usepackage{algorithmic}
    \usepackage{enumitem}

   \newcommand{\bi}[1]{\textbf{\textit{#1}}}
   \defcitealias{wang_particle--cell_2011}{WMC2011}


\shorttitle{The Convergence of Cosmological PIC Schemes}
\shortauthors{Myers, Colella, \& Van Straalen}
\begin{document}
\title{The Convergence of Particle-in-Cell Schemes for Cosmological Dark Matter Simulations}
\author{Andrew Myers, 
            Phillip Colella,
            and Brian Van Straalen}

\affil{Lawrence Berkeley National Laboratory, 1 Cyclotron Road, Berkeley, CA, 94720, USA}
\email{ATMyers@lbl.gov}
         
\begin{abstract}
 	Particle methods are a ubiquitous tool for solving the Vlasov-Poisson equation in comoving coordinates, which is used to model the gravitational evolution of dark matter in an expanding universe. However, these methods are known to produce poor results on idealized test problems, particularly at late times, after the particle trajectories have crossed. To investigate this, we have performed a series of one- and two-dimensional ``Zel'dovich Pancake" calculations using the popular Particle-in-Cell (PIC) method. We find that PIC can indeed converge on these problems provided the following modifications are made. The first modification is to regularize the singular initial distribution function by introducing a small but finite artificial velocity dispersion. This process is analogous to artificial viscosity in compressible gas dynamics, and, as with artificial viscosity, the amount of regularization can be tailored so that its effect outside of a well-defined region - in this case, the high-density caustics - is small. The second modification is the introduction of a particle remapping procedure that periodically re-expresses the dark matter distribution function using a new set of particles. We describe a remapping algorithm that is third-order accurate and adaptive in phase space. This procedure prevents the accumulation of numerical errors in integrating the particle trajectories from growing large enough to significantly degrade the solution. Once both of these changes are made, PIC converges at second order on the Zel'dovich Pancake problem, even at late times, after many caustics have formed. Furthermore, the resulting scheme does not suffer from the unphysical, small-scale ``clumping"  phenomenon known to occur on the Pancake problem when the perturbation wave vector is not aligned with one of the Cartesian coordinate axes. 
 \end{abstract}
 
 
\section{Introduction}

    Cosmological dark matter (DM) simulations follow the evolution of a
self-gravitating, collisionless fluid in a coordinate system that
expands along with the universe. This fluid can either be treated in
isolation or coupled to a second baryonic fluid that is subject to its
own set of physical processes. Such simulations cover a wide range of
length scales, from cosmic-scale  ($\sim$ 10
Gpc) calculations that cover a
representative volume of the observable universe \citep[e.g.][]{deus_2012}, down to galactic scale ($\sim$ 10 kpc) calculations
that probe the substructure of the DM fluid within Milky-Way sized
($\sim$ \(10^{12}\) \(M_{\odot}\)) halos \citep[e.g.][]{kuhlen_2008}. The results of
these simulations are used to guide, interpret, or calibrate virtually
all current and planned observational efforts to study the nature of dark matter and dark
energy, including direct and indirect dark matter detection, galaxy
redshift surveys, and weak lensing studies \citep{kuhlen_review}.

The evolution of the DM fluid is governed by the Vlasov-Poisson equation
in comoving coordinates. While it is possible to solve this set of
equations using grid-based methods in phase space \citep{yoshikawa_direct_2013}, 
this approach is not common in practice due to the computational expense of
working in high-dimensional spaces. Typically, particle methods are used instead. 
These methods are based on a Lagrangian description of the Vlasov-Poisson system,
in which the initial dark matter distribution function is 
sampled by a finite set of particles. The particle positions and velocities are then advanced in time using the appropriate 
equations of motion derived from the Vlasov-Poisson equation, combined with some sort of scheme
for computing the forces at the particle positions. This force calculation can be carried out in a variety of ways. In PIC methods, for example, an Eulerian representation of the density is constructed from the particle positions by  deposition and the Poisson equation for the gravitational potential is solved on the resulting mesh points. An example of a cosmological DM code that uses pure PIC for its force solve is MC\(^2\) \citep{heitmann_robustness_2005};
other popular choices are tree codes, e.g. pdkgrav2
\citep{stadel_cosmological_2001} and 2HOT \citep{warren_2hot:_2013},
adaptive PIC codes, e.g. ART \citep{kravtsov_adaptive_1997}, 
RAMSES \citep{teyssier_cosmological_2002} and Nyx \citep{nyx}, or combinations of the two, e.g. GADGET2 \citep{springel_cosmological_2005} and HACC
\citep{habib_universe_2012}. While the details of the force solves
differ, all of these codes share the underlying particle discretization of the 
Vlasov-Poisson system. Such calculations have now
been run with hundreds of millions \citep{alimi_deus_2012} to over a
trillion \citep{ishiyama_4.45_2012, skillman_dark_2014} particles.

    The accuracy of cosmological observations has advanced tremendously
over the past few decades, and with the advent of next-generation
instruments like the Large Synoptic Survey Telescope \footnote{http://www.lsst.org/lsst} in sight, the
prospect of measuring the basic cosmological parameters of the universe
to better than \(1\%\) accuracy has become conceivable \citep[e.g.][]{heitmann_coyote_2009, heitmann_coyote_2010}. However, interpreting these observations will require theoretical predictions of similar or better accuracy, and it is well-known that standard particle techniques produce poor results
on some idealized test problems. A particularly striking example is the
off-axis plane wave collapse problem of \cite{melott_demonstrating_1997}. \citeauthor{melott_demonstrating_1997} set up a plane wave perturbation that was arbitrarily
aligned relative to the Cartesian coordinate axes and traced its
evolution into the non-linear regime using several different particle-based numerical schemes. All of the schemes failed to converge to the correct solution when sparse particle counts were used. Instead of respecting the symmetry of the problem setup, non-physical clumping was observed in directions perpendicular to the axis of the perturbation. This result has since been reproduced by several researchers \citep[]{heitmann_robustness_2005, hahn_new_2013}.

The inability of cosmological dark matter simulations to converge on this problem is troubling. It is not clear to what extent these problems manifest themselves in more realistic calculations.  Code comparisons involving more realistic initial conditions generally show encouraging agreement on physical observables for dark-matter only
problems \citep[e.g.][]{heitmann_robustness_2005, heitmann_cosmic_2008, kim_agora_2014}. However, all of these codes solve the Vlasov-Poisson system using particle methods, differing mainly in the
way the force solve is carried out. Thus, these comparisons cannot rule
out the hypothesis that common assumptions are giving rise to
common errors. A related issue is the small-scale artificial
fragmentation observed in particle simulations of warm dark matter \citep{wang_discreteness_2007}. This fragmentation may be related to the ``clumping" effect observed in the off-axis pancake problem, but this needs to be demonstrated explicitly. Additionally, as pointed out by \cite{hahn_new_2013}, we cannot be sure that cold dark matter simulations
do not suffer from the same problems, but are simply hidden by the presence of initial 
power on small scales. 

    In light of this, a particle method that can be rigorously shown to converge - even on a simple test problem like the Zel'dovich Pancake - is desirable. The problem of improving the accuracy of particle methods on cosmology calculations has been considered by \cite{hahn_new_2013}, who presented an improved PIC scheme that exploits the continuity of the infinitesimally thin DM sheet in phase space. We take an alternative approach, based on the convergence theory for PIC methods from \citet[][hereafter WMC2011]{wang_particle--cell_2011}, which focused on the application of PIC to electrostatic plasmas. That theory shows that PIC methods can converge on the Vlasov-Poisson problem provided that 1) the initial particle spacing is less than or equal to the cell spacing of the Poisson mesh, and 2) a phase-space remapping procedure is applied to the particles that prevents the accumulation of numerical error in the particle trajectories from overwhelming the solution. However, this theory assumes that the initial distribution function is finite, and hence does not apply to the cold dark matter version of the Vlasov-Poisson problem, where the initial distribution function is usually taken to be a delta-function in velocity space. Thus, in order to use this theory to guide our investigation of the cold dark matter problem, we will regularize the initial conditions so that the convergence theory in \citetalias{wang_particle--cell_2011} applies. A natural way to do so is to introduce a finite, artificial velocity dispersion $\sigma_i$ to the initial DM distribution function. Once solutions to these artificially warm versions of the cold dark matter problem are obtained, they can used to be used to obtain solutions to the perfectly cold problem by taking the limit $\sigma_i \rightarrow 0$.
    
 An instructive analogy is to the use of an artificial viscosity in shock-capturing schemes for
compressible gas dynamics. For Reynolds numbers characteristic of
supersonic flows in astrophysics, the molecular mean free path of the
gas is tiny compared to the other length scales of interest, and thus
the effects of viscosity should be negligible. However, neglecting the
viscous terms entirely leads to numerical instabilities that can
contaminate the solution far from any shocks. A common solution is to
introduce an artificially large numerical viscosity that smooths out the
solution near a shock front over a few grid cells. Away from the shock,
the integrity of the solution is preserved. Likewise, in structure
formation, the present-day microscopic velocity dispersion of dark
matter particles is negligibly tiny compared to the bulk velocities that
arise from gravitational collapse. However, ignoring the velocity dispersion
completely leads to resolution-dependent results at the caustic locations, and could potentially cause other problems as well. Using an artificially large initial
velocity dispersion smooths out the dark matter caustics over a length scale that is
numerically resolvable, preserving the qualitative features
of the dark matter density distribution without affecting the solution
outside of the caustics.
    
The second modification suggested by the error analysis of \citetalias{wang_particle--cell_2011} is to periodically \emph{remap} 
the DM particles in phase space. This addresses a downside of particle methods, which is that numerical errors made in integrating 
the particle trajectories tend to compound with each successive time step. In electrostatic PIC, this phenomenon manifests itself as an exponentially growing term in the stability error for the electric field \citepalias[][Equation 3.12]{wang_particle--cell_2011}. This term eventually overwhelms the solution as the simulation evolves. To mitigate this,
the distribution function must be periodically re-expressed using a
new set of particles before integration errors have a chance to
accumulate. This procedure is crucial for obtaining quality solutions
to plasma problems for long time integrations. Note that this procedure requires the above regularization of the initial data, in that we must be able to generate a well-defined distribution function in phase space corresponding to a given particle distribution.

    In this paper, we show that this approach improves the accuracy of PIC on the Zel'dovich Pancake test problem. The structure of the paper is as follows. We begin by reviewing 
the cosmological Vlasov-Poisson system and the standard
PIC algorithm for solving it in Section \ref{sec:methods}. In Section \ref{sec:singular}, we perform a series of convergence studies that test PIC's ability to track the gravitational collapse of a single, plane-wave perturbation to late times. We perform two versions of this test, first with only one spatial dimension, and second in two spatial dimensions, with the axis of the perturbation misaligned with the Cartesian coordinate axes. We confirm that standard PIC with singular initial data converges poorly at late times in 1D, and that it suffers from unphysical clumping on the 2D, oblique problem, regardless of the number of particles per Poisson cell employed. Next, in Section \ref{sec:regularized}, we consider the same numerical method, using initial data that has been regularized via the introduction of a finite artificial velocity dispersion. We show that, given finite initial conditions, the standard PIC technique does converge at 2nd order in 1D. We also show that, in the limit that the artificial velocity dispersion goes to zero, we recover the solution to the problem with perfectly cold initial conditions. However, we find that the introduction of regularization does not, by itself, solve the problem with artificial clumping on the 2D, oblique problem. Finally, in Section \ref{sec:remapping}, we introduce particle remapping. We perform this remapping on an adaptive set of grids that automatically tracks the evolution of the dark matter distribution function in phase space as the universe expands. Our results using both regularization and remapping are shown in Section \ref{sec:remapped_results}. We show that, when remapping is employed, 2nd order convergence results are still obtained in 1D, and the solution is not affected by the artificial clumping present in other cases. We summarize our conclusions in Section \ref{sec:conclusions}.
        
    \section{Equations and Algorithms}
    \label{sec:methods}
    \subsection{The Vlasov-Poisson system}
    We work with a non-dimensional form of the Vlasov-Poisson equation in comoving coordinates:
    
\begin{equation}
\label{vlasov-poisson}
\frac{\partial f}{\partial t} = - \frac{\bi{v}}{a} \cdot  \frac{\partial f}{\partial \bi{x}} + \left[ \left( \frac{\dot{a}}{a} \right) \bi{v} + \frac{1}{a} \nabla \phi \right] \cdot \frac{\partial f}{\partial \bi{v}}.  
\end{equation} Here, $f(\bi{x}, \bi{v})$ is the dark matter distribution function in $2 D$-dimensional phase space \((\bi{x}, \bi{v}) \in \mathbb{R}^{D} \times \mathbb{R}^{D}\), where $D$ is the number of spatial dimensions under consideration. $\bi{x}$ and $\bi{v}$ are the usual comoving position and peculiar velocity coordinates \citep{peebles_principles_1993}, related to the proper position coordinate $\bi{r}$ by 

\begin{align}
\bi{x} &= a(t)^{-1} \bi{r} \nonumber \\
\bi{v} &= a(t) \dot{\bi{x}}. 
\end{align}

The expansion of the universe is described by the time-dependent scale factor $a(t)$. In general, the time evolution of \(a(t)\) is determined by the assumed
cosmological parameters. In this paper, we specialize to a flat,
critical-density, matter-only universe, so that

 \begin{equation}
a(t) = \left( \frac{3 t}{2} \right)^{2/3}.
\end{equation}

The gravitational potential $\phi$ obeys the Poisson equation:
\begin{equation}
\nabla^2 \phi = \frac{3}{2 a} \left( \rho - 1 \right),
\end{equation} where $\rho$ is the comoving dark matter density. We complete the specification of the
problem by adopting periodic boundary conditions on the domain
\(\bi{x} \in (0, 1)^D\) and supplying appropriate initial conditions, $f(\bi{x}, \bi{v}, t = t_{\text{ini}})$.

\subsection{Particle Discretization}

In particle methods for numerically solving Equation \ref{vlasov-poisson}, the initial distribution function is discretized by a set of Lagrangian particles, $\mathbb{P}$. Given $f(\bi{x}, \bi{v}, t_{\text{ini}})$, we must choose a set of particles with initial positions $\bi{x}_p^i$, velocities $\bi{v}_p^i$, and masses $m_p$ such that 

\begin{equation}
    \label{discrete_f}
f(\bi{x}, \bi{v}, t_{\text{ini}}) \approx \sum_{p \in \mathbb{P}} m_p \delta \left(\bi{x} - \bi{x}_p^i \right) \delta \left(\bi{v} - \bi{v}_p^i \right).
\end{equation}

From these initial coordinates, the particles evolve according to the
appropriate equations of motion. The trajectory
\(\left( \bi{x}_p(t), \bi{v}_p(t) \right)\) of
each particle can be obtained by substituting Equation \ref{discrete_f} into Equation \ref{vlasov-poisson}. The result is the following system of ODEs:

    \begin{align}
    \label{eq:particle_eom}
\frac{d m_p}{dt} &= 0 \nonumber \\
\frac{d \bi{x}_p}{dt} &= \frac{1}{a} \bi{v}_p \nonumber \\
\frac{d \bi{v}_p}{dt} &= - \frac{\dot{a}}{a} \bi{v}_p + \frac{1}{a} \bi{g}_p,
\end{align}

where \( \bi{g}_p\) is the acceleration on particle \(p\)
induced by the dark matter distribution. It is important to note that these particles have nothing to do with the actual dark matter particles whose statistics are described by the Vlasov-Poisson equation; they are simply a set of interpolating points from which the distribution function can be recovered. 

\subsection{The PIC Update}
\label{sec:pic_update}
   
\begin{figure}
\label{control_pic}
\epsscale{1.2}
\plotone{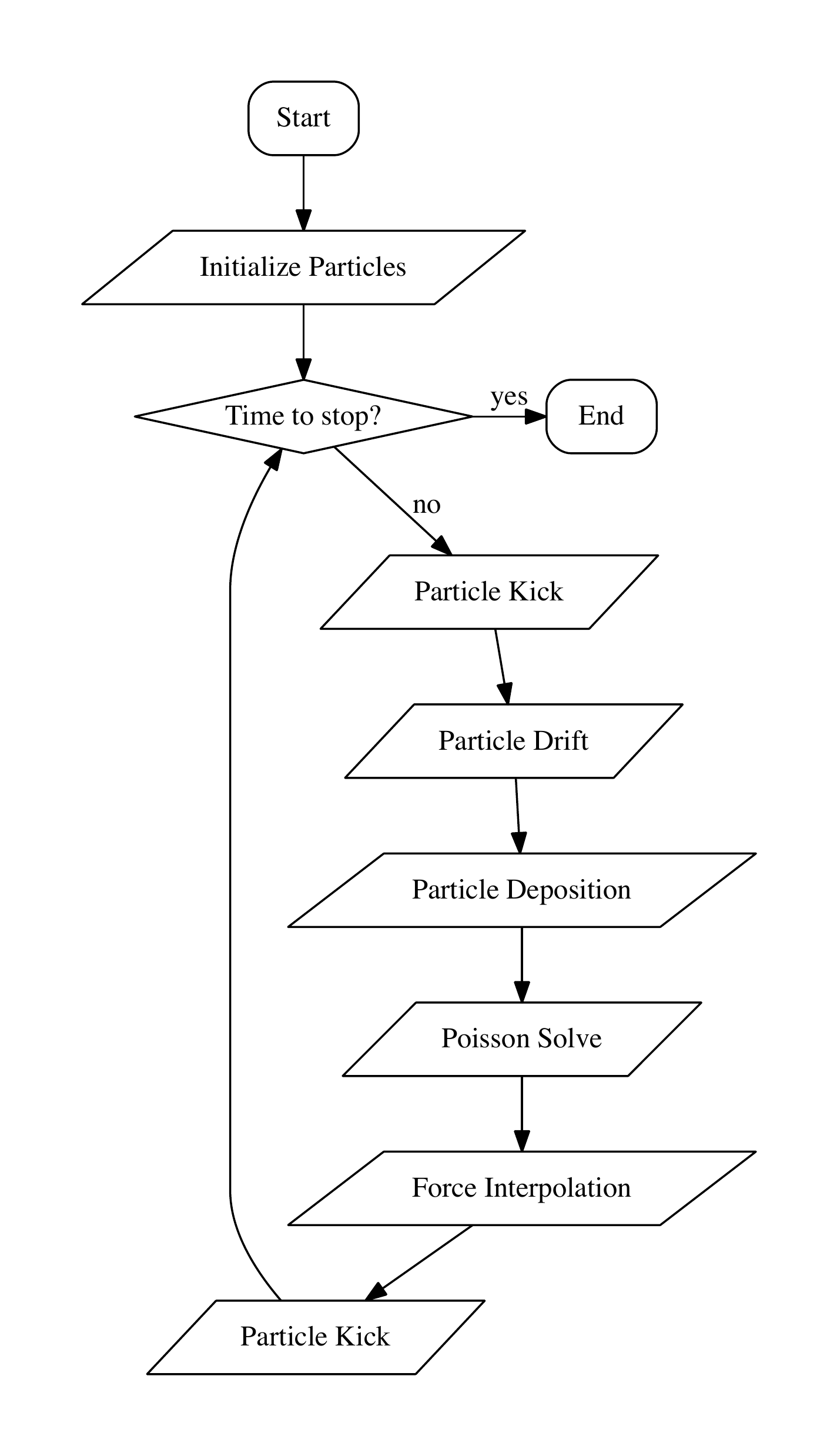}
\caption{The overall control flow of a PIC calculation}
\end{figure}
   
    Once the initial particles have been generated, we advance their positions and velocities using a standard Particle-in-Cell technique \citep{hockney_computer_1981}. We write the particle comoving positions and peculiar velocities at time $t^n$ as $\bi{x}_p^n$ and $\bi{v}_p^n$, where $n$ is the time step number. We also assume that we know the gravitational field at the particle positions $\bi{g}_p^n$ at the same time index, either by performing an initial Poisson solve or by storing the result from the previous time step.  Given that information, we advance the particles to time $t^{n+1}$ as follows:

\textbf{Particle Kick.} For our time integration scheme, we use the second-order and sympleptic Kick-Drift-Kick sequence from \cite{miniati_block_2007}.  Note that this scheme is 
self-starting in the sense that it does not require any information from time
steps prior to \(n\) to complete the update - an important feature given our use of 
remapping below. We begin by updating the particle velocities to the half-time index $n+1/2$: 
\begin{equation}
\bi{v}_p^{n+1/2} = \frac{a^n}{a^{n+1/2}} \bi{v}_p^n + \frac{1}{a^{n+1/2}} \bi{g}_p^{n}\frac{\Delta t}{2}.
\end{equation} Here, \(\Delta t\) is the time step, and \(a^n\) and \(a^{n+1/2}\) are the
expansion factors computed at times \(t^{n}\) and \(t^{n+1/2}\). 

\textbf{Particle Drift.} Next, the particle positions are advanced using the half-time velocity:
\begin{equation}
\bi{x}_p^{n+1} = \bi{x}_p^{n} + \frac{1}{a^{n+1/2}} \bi{v}_p^{n+1/2} \Delta t. 
\end{equation} The final update to the particle velocities cannot be completed until we compute the new forces at time index $n+1$.  

\textbf{Particle Deposition.} To do so, we compute the density at the mesh points \(\bi{x}_{\bi{i}} = \left( \bi{i} + 1/2 \right) \Delta \bi{x}\), where \(\bi{i} \in \mathbb{Z}^D\) are the cell indices:
\begin{equation}
\label{eq:particle_deposit}
\rho_{\bi{i}}^{n+1} = \sum_p \left( \frac{m_p}{V_i} \right) \bi{W}_{\bf{CIC}} \left( \frac{\bi{x}_{i} - \bi{x}_p^{n+1}}{\Delta x} \right).
\end{equation} Here, $V_i = {\Delta x}^D$ is the volume of cell $i$ and $\bi{W}_{\bf{CIC}}(\bi{x})$ is the $D$-dimensional cloud-in-cell (CIC) interpolating function: 
\begin{equation}
\bi{W}_{\bf{CIC}} \left(\bi{x} \right) = \prod_{d = 1}^D W_{\text{CIC}} \left( x_d \right),
\end{equation}
\begin{equation}
   W_{\text{CIC}}(x) = \left\{
     \begin{array}{lr}
       1 - \lvert x \rvert, & 0 < \lvert x \rvert < 1, \\
       0 & \text{otherwise.}
     \end{array}
   \right.
\end{equation} Note that in general, we do not use the same mesh spacing for the particle discretization and the Poisson mesh, i.e. $\Delta x \ne h_x$.

\textbf{Poisson Solve.} The next step is to solve the Poisson equation for the gravitational potential at the same grid points on which the density is defined. We discretize the Laplacian operator using the standard \(2 D\)+\(1\) point centered difference approximation:
\begin{equation}
- \sum_{d=1}^D \frac{\phi_{\bi{i} + \bi{e}^d}^{n+1} - 2\phi_{\bi{i}}^{n+1} + \phi_{\bi{i} - \bi{e}^d}^{n+1}}{{\Delta  x}^2} = \frac{3}{2 a^{n+1}} \left( \rho_{\bi{i}}^{n+1} - 1 \right).
\end{equation} To solve the resulting linear system, we use the geometric
multigrid package in Chombo \citep{chombo_design}, using Gauss-Seidel with Red-Black ordering
as the smoother. We set the solver tolerance to \(10^{-9}\) - small enough its precise value does not affect our convergence results below.  After iterating to convergence, the gravitational field
\(\bi{g} = -\nabla \phi\) can be computed at the same grid points as
\begin{equation}
\label{PICForce}
\bi{g}_{\bi{i}}^{n+1} = - \frac{\phi_{\bi{i} + \bi{e}^d}^{n+1} - \phi_{\bi{i}^{n+1} - \bi{e}^d}^{n+1}}{2 \Delta x}.
\end{equation}
\textbf{Force Interpolation.} Next, we interpolate the field back to the particle positions using the same interpolating function as in the deposition step:
\begin{equation}
\bi{g}_p^{n+1} = \sum_{\bi{j}} \bi{g}_{\bi{i}} V_i \bi{W}_{\bf{CIC}} \left( \frac{\bi{x}_j -\bi{x}_p^{n+1}}{\Delta x} \right).
\end{equation}

\textbf{Particle Kick.} Finally, we complete the update to the particle positions as:
\begin{equation}
\bi{v}_p^{n+1} = \frac{a^{n+1/2}}{a^{n+1}} \bi{v}_p^{n+1/2} + \frac{1}{a^{n+1}} \bi{g}_p^{n+1}\frac{\Delta t}{2}.
\end{equation}

    \subsection{Time stepping strategy}
    \label{sec:time_stepping}
    
    In this paper, the size of the time step $\Delta t$ is controlled by two factors. First, we limit the factor by which the background can expand in a single time step. The time step associated with the expansion factor is:
    \begin{equation}
    \label{eq:expansion_time_step}
    \Delta t_{\text{exp}} = C_{\text{exp}} \left( \frac{a}{\dot a} \right),
    \end{equation} where $C_{\text{exp}}$ is a tunable parameter. 
    
    Second, we enforce a Courant-Friedrichs-Lewy (CFL) -type constraint, which limits the distance that a particle can travel in a single time step to be some fraction of $\Delta x$. The time step associated with this constraint is:
    \begin{equation}
    \Delta t _{\text{part}} = C_{\text{part}} \frac{\Delta x}{\max( | \bi{v}_p |)}.
    \end{equation} In this paper, we take $C_{\text{part}} = 0.5$. We use the smaller of these two time steps to advance the simulation:
    \begin{equation}
    \Delta t = \min ( \Delta t_{\text{exp}}, \Delta t _{\text{part}} ).
    \end{equation} This approach to time stepping is quite similar to that in, e.g., \cite{nyx}. In the runs presented in this paper, the expansion of the background typically controls $\Delta t$ at early times, while the particle CFL constraint controls it at late times.

Given an initial set of particles, the above equations describe how to advance the system to an arbitrary later time. The overall control flow of the PIC procedure is illustrated in Figure \ref{control_pic}. We have implemented the above algorithm, as well as the remapping procedure described in Section \ref{sec:remapping}, using the Chombo\footnote{https://commons.lbl.gov/display/chombo} software framework for partial differential equations, using standard message passing with block-structured domain decomposition for parallelization. This code, as well as our data analysis scripts, are available online\footnote{https://bitbucket.org/atmyers/cosmologicalpic}. 

\section{The Zel'dovich Pancake}
\label{sec:singular}
We now study the convergence of the PIC method for plane-wave initial conditions.
The gravitational collapse of a plane-wave perturbation in an expanding background, 
sometimes called the ``Zel'dovich Pancake'' problem
because of the flattened structures it produces, is a common test case
for cosmological codes \citep[e.g.][]{bryan1995, kravtsov_adaptive_1997, miniati_block_2007, gizmo} for several reasons. At early times (before any
matter parcels cross) an exact solution exists in 1D, making the problem
valuable for code validation. After the first shell crossing, the
problem still provides a valuable test on a code's ability to deal with strong
density contrasts and poorly resolved features. Finally, while the
problem is clearly idealized, it still physically relevant, as the collapse of a single Fourier mode forms the basis for more complex structure formation calculations. 

\subsection{Initial conditions}

The initial distribution function is taken to be an infinitesimally thin sheet in
phase space:
\begin{equation}
\label{cold_f}
f(\bi{x}, \bi{v}, t_{\text{ini}}) = \rho_z \left( \bi{x}, t_{\text{ini}} \right) \delta \left( \bi{v} - \bi{v}_z \left( \bi{x}, t_{\text{ini}} \right) \right).
\end{equation} 

    The initial density \(\rho_z \left( \bi{x}, t_{\text{ini}} \right)\)
and velocity \(\bi{v}_z \left( \bi{x}, t_{\text{ini}} \right)\) are then computed using the Zel'dovich approximation. For
example, if the matter parcels of the unperturbed DM fluid are labelled
by their comoving Lagrangian coordinates \(\bi{q}\), and we apply a
sinusoidal perturbation of the form
    \begin{equation}
\bi{S}(\bi{q}) = \frac{5 A}{2} \sin(\bi{k} \cdot \bi{q}) \hat{\bi{k}},
\end{equation}

    where \(A\) is the amplitude and \(\bi{k}\) the wavenumber, then (in
our adopted Einstein-de Sitter cosmology) the comoving position and
peculiar velocity of the fluid at time \(t\) are
    \begin{align}
    \label{Zel'dovich}
\bi{x}(t) &= \bi{q} + a A \sin(\bi{k} \cdot \bi{q}) \hat{\bi{k}} \\
\bi{v}(t) &= a \dot{a} A \sin(\bi{k} \cdot \bi{q}) \hat{\bi{k}}.
\end{align}

    The corresponding density is
    \begin{equation}
\rho(\bi{q}, t) = \frac{1}{1 + a A | \bi{k} | \cos(\bi{k} \cdot \bi{q})},
\end{equation}

which can be converted to Eulerian coordinates \(\bi{x}\) using Equation \ref{Zel'dovich}.

    The above equations are valid until the first caustic forms, which
happens at the expansion factor
\begin{equation}
a_{\text{caustic}}(t) = \frac{1}{A \lvert \bi{k} \rvert}.
\end{equation}

At that point, the velocity becomes multi-valued and the density
diverges at the positions corresponding to
\(\cos(\bi{k} \cdot \bi{q}) = 1\). Prior to that time, we can use
the above equation to set initial conditions. In particular, we represent the above initial conditions with a collection 
of \(N_p\) equal-mass particles
labelled by the Lagrangian coordinates
\begin{equation}
\bi{q}_p = \left( \bi{i} + 1/2 \right) h_x, 
\end{equation} where \(h_x = 1 / N_p\) is the initial particle spacing 
(which is not necessarily the same as the mesh spacing for the Poisson solve, $ \Delta x$),
\(\bi{i} \in \mathbb{Z}^D\), and $m_p = 1 /
N_p$ is the particle mass. We start the integration at time \(t_{\text{ini}}\),
corresponding to expansion factor \(a_{\text{ini}}\). At that time, the
Eulerian positions and velocities of the particles are

\begin{align}
\bi{x}_p^i &= \bi{q}_p + a_{\text{ini}} A \sin(\bi{k} \cdot \bi{q}_p) \hat{\bi{k}} \\
\bi{v}_p^i &= a_{\text{ini}} \dot{a}_{\text{ini}} A \sin(\bi{k} \cdot \bi{q}_p) \hat{\bi{k}}.
\end{align}

    Note that after applying a perturbation, neither the particle positions
nor the velocities are laid out on a Cartesian grid. This procedure, in which 
the Zel'dovich approximation is used to construct initial particle positions and velocities 
and the initial distribution function is perfectly cold, is standard in cosmological dark matter 
simulations \citep[e.g.][]{hahn_multi-scale_2011}.

\subsection{1D results}
\label{sec:singular_results_1D}

\begin{deluxetable}{ccc}
\tablecaption{\label{1D_table_sin} Summary of parameters for the singular, 1D runs}
\tablewidth{0pt}
\tablehead{
\colhead{$N_{\text{cells}}$} & \colhead{$n_x$} & \colhead{$C_{\text{exp}}$}
}
\startdata
    $256$ & 128 & $1.0 \times 10^{-2}$  \\
    $512$ & 128 & $5.0 \times 10^{-3}$ \\
    $1024$ & 128 & $2.5 \times 10^{-3}$ \\
    $2048$ & 128 & $1.25 \times 10^{-2}$ \\
\enddata
\end{deluxetable}

\begin{figure*}
\epsscale{1.2}
\plotone{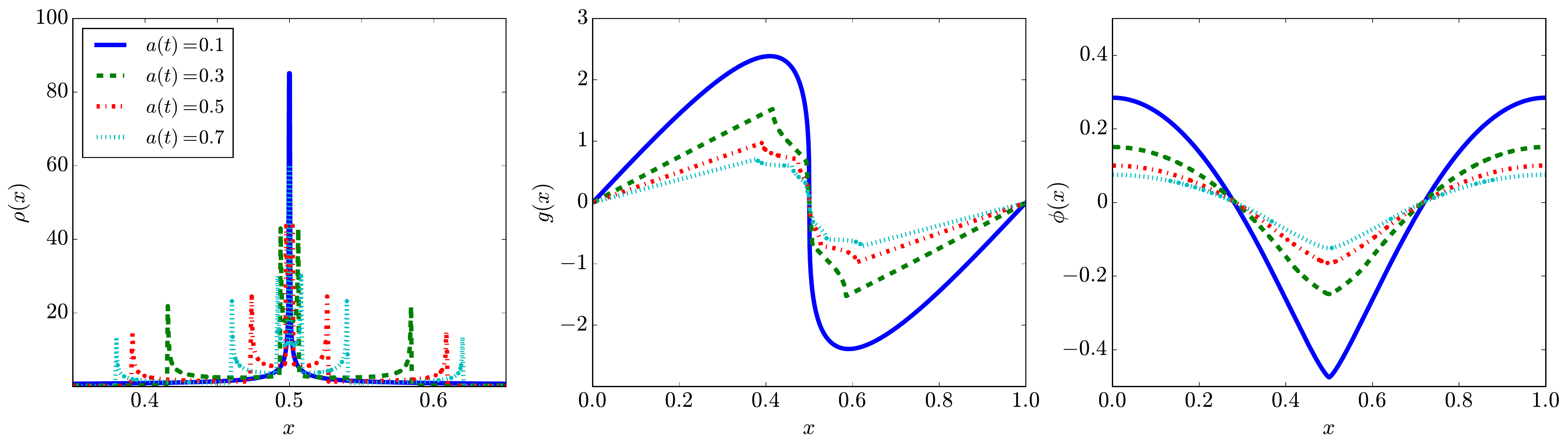}
\caption{The time evolution of the solution to the 1D, singular problem. The left panel shows the density, the middle panel the gravitational field, and the right panel the potential. The plotted curves are from the run with $N_{\text{cells}} = 2048$. For the gravitational field and the potential, the $N_{\text{cells}} = 1024$ run would be indistinguishable from the plotted solutions at the scales shown. The solutions for the density, on the other hand, are not converged; see Figure \ref{1D_density_converge_singular}. The different lines correspond to the results at different expansion factors: solid blue line - $a(t) = 0.1$; dashed green line - $a(t) = 0.3$; dashed-dotted red line - $a(t) = 0.5$; dotted cyan line - $a(t) = 0.7$.  }
\label{evolution_singular}
\end{figure*}

We begin by performing a resolution study on a 1D problem setup. This is the easiest test for the method to pass, since some numerical instabilities may only be possible in multi-dimensional problems. However, this test will help us verify that we have implemented the above scheme correctly, and it may also provide clues about the behavior of the higher dimensional case. 

\begin{figure}
\epsscale{1.2}
\plotone{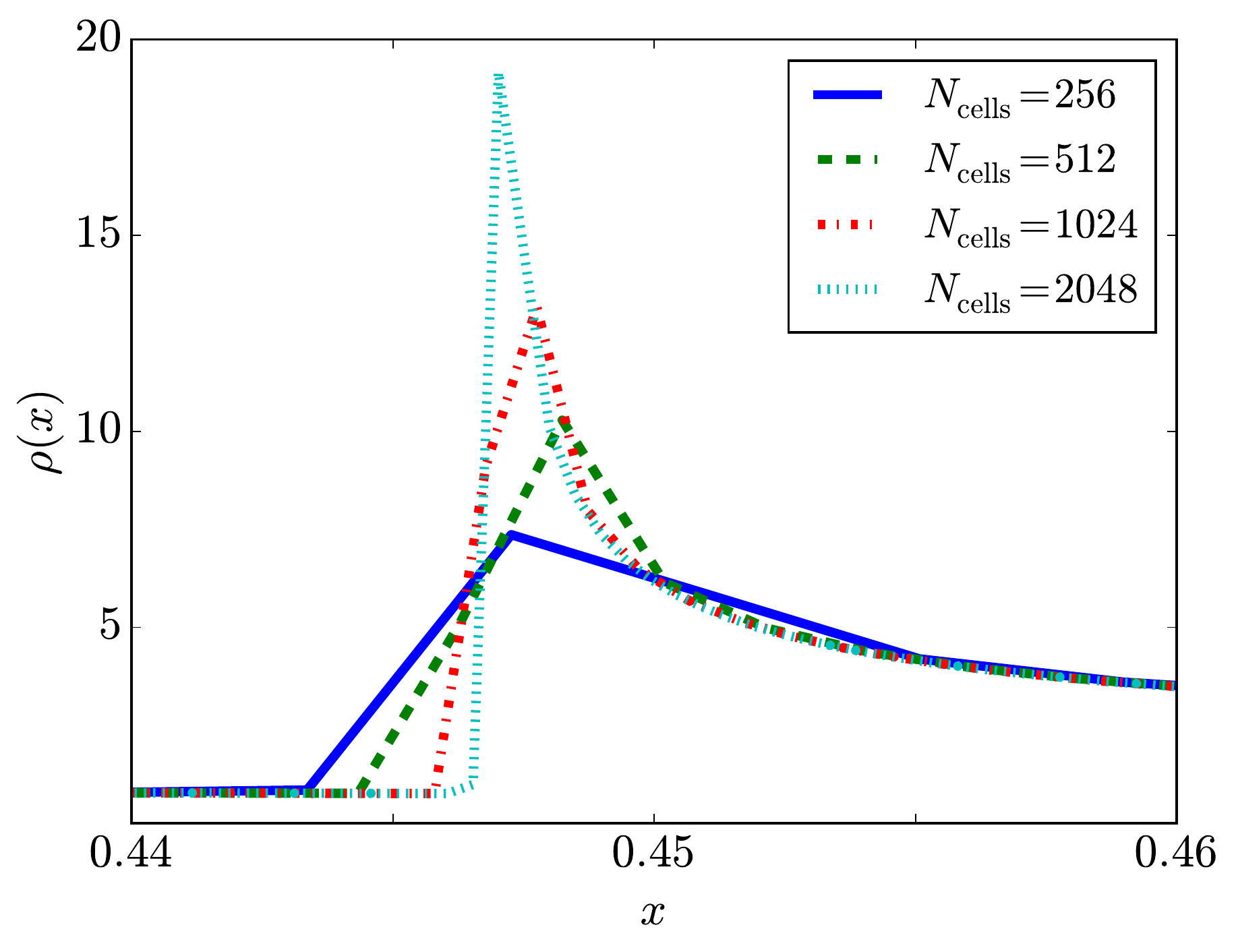}
\caption{A zoomed-in view of one of the outer density caustics in the 1D, singular problem, taken at $a(t) = 1$. The different colors correspond to different resolutions: solid blue line - $N_{\text{cells}} = 256$; dashed green line - $N_{\text{cells}} = 512$; dashed-dotted red line - $N_{\text{cells}} = 1024$; dotted cyan line - $N_{\text{cells}} = 2048$. }
\label{1D_density_converge_singular}
\end{figure}

For these runs, we set the perturbation wavenumber to $2 \pi$, the fundamental mode of the computational box. We initialize the problem at $a_{\text{ini}} = 1 / 200$ and choose the perturbation amplitude so that shell crossing occurs at $a_{\text{caustic}} = 0.1$. We then evolve the simulations until $a_{\text{stop}} = 1$, using the scheme described in Section \ref{sec:pic_update}. The simulation results are dumped out every $\Delta a = 0.01$, i.e. at $a = 0.01$, $0.02$, ..., $0.99$, $1.00$. We fix the number of particles per Poisson cell, $n_x$, where $n_x = N_p / N_{\text{cells}}$, to $128$, and vary the number of Poisson cells in the problem domain $N_{\text{cells}} = 1 / {\Delta x}$ over the range \{$256$, $512$, $1024$, $2048$\}. Every time we decrease the mesh spacing, we also decrease the parameter $C_{\text{exp}}$ by a factor of 2. These parameters are summarized in Table \ref{1D_table_sin}. Note that our use of 128 particles per cell is something of a best-case scenario; most calculations in cosmology use significantly fewer particles. The resulting solutions are shown at selected times in Figure \ref{evolution_singular}. As expected, at $a(t) = 0.1$ the particle trajectories cross each other and a single dark matter caustic forms. As the particles continue to evolve in phase space, secondary and tertiary caustics form. Each caustic is marked by a peak in the density profile and cusp in the gravitational field. The density peaks in the dark matter caustics are clearly not converged, as is to be expected with singular initial data (see Figure \ref{1D_density_converge_singular}). Visually, however, the gravitational field and the potential appear to approach well-defined solutions. 

To verify this, we compute the convergence rates for the density, gravitational field, and potential. While an analytic solution to the Zel'dovich Pancake problem exists prior to $a_{\text{caustic}}$, we are also interested in the convergence behavior at late times, and it is convenient to have an error metric that applies to both $a < a_{\text{caustic}}$ and $a > a_{\text{caustic}}$. Additionally, as we will later want to apply our metric to regularized and remapped runs, we cannot compare particle quantities directly. Instead, we must focus on the grid-defined quantities used in the PIC update: $\bi{g}_{\bi{i}}$, $\rho_{\bi{i}}$, and $\phi_{\bi{i}}$.  We therefore use Richardson extrapolation to compute our error estimates. If $Q^h$ is one of these grid quantities, and $Q^{2 h}$ is the same quantity computed with $\emph{all}$ the discrete elements ($\Delta x$, $h_x$, $\Delta t$, $C_{\text{exp}}$) coarsened by a factor of 2, then we define the relative solution error as
\begin{equation}
e^h = \lvert Q^h  - Q^{2 h} \rvert.
\end{equation} To compare solutions with different numbers of grid points, we average the finer solution down to the resolution of the coarse solution. The order $q$ of the method is then
\begin{equation}
q = \log_2 \left( \frac{\lvert \lvert e^{2 h} \rvert \rvert }{\lvert \lvert e^{h} \rvert \rvert } \right),
\end{equation} where $\lvert \lvert x \rvert \rvert$ is one of the $L_1$, $L_2$, or $L_{\infty}$ norms of $x$. In what follows, we use the symbols $q_{\rho}$, $q_{g}$, and $q_{\phi}$ to denote the order of convergence in the density, gravitational field, and potential, respectively. 

\begin{figure*}
\epsscale{1.2}
\plotone{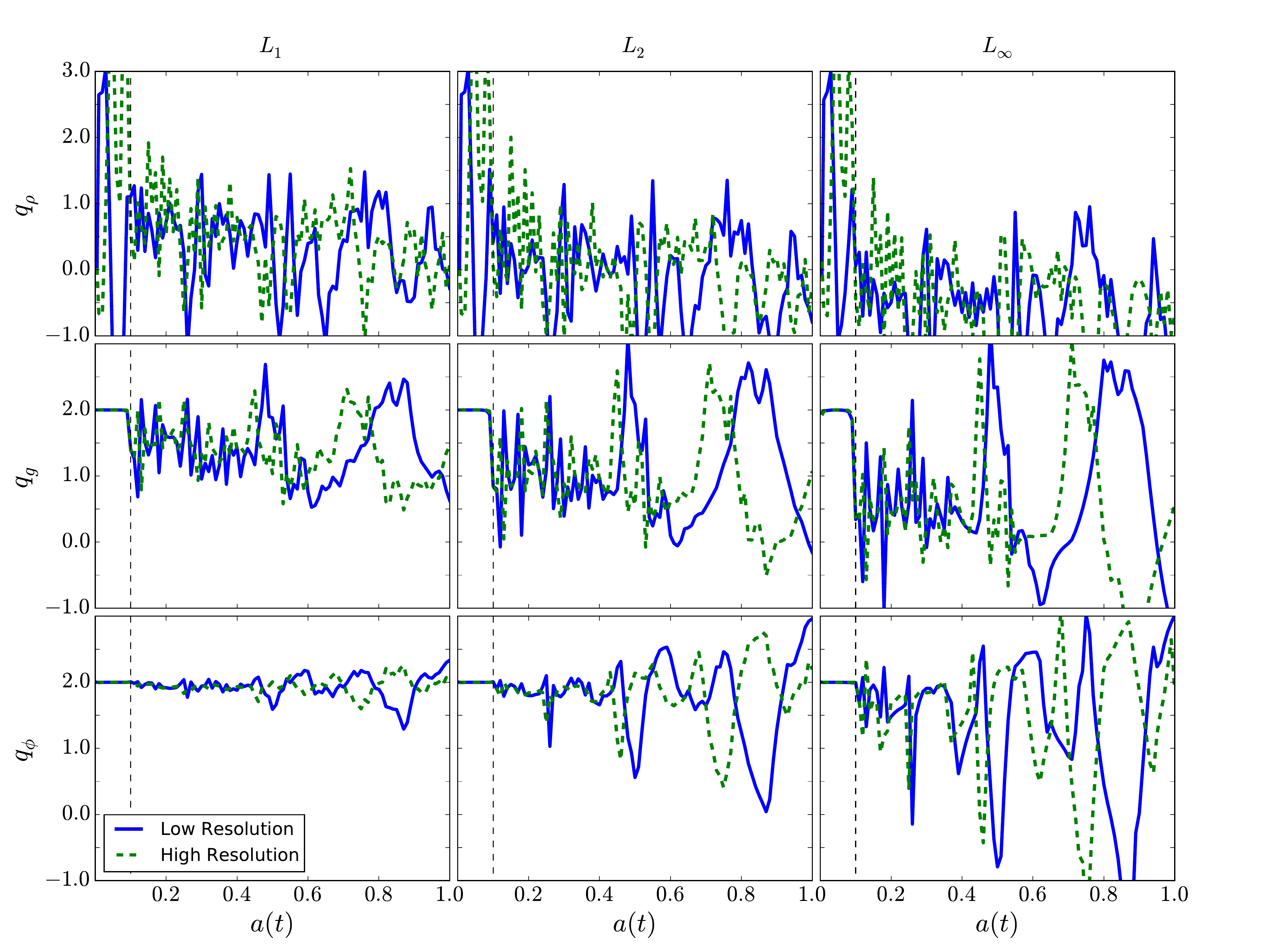}
\caption{Convergence results for the singular version of the 1D pancake problem. The plotted curves show the Richardson-estimated order versus the expansion factor $a(t)$. For the solid blue line, the order was estimated using the runs with $N_{\text{cells}} = 256$, $512,$ and $1024$, while the dashed green line shows the same quantity for resolutions $N_{\text{cells}} = 512$, $1024,$ and $2048$. The top row shows the rates computed for the density, the middle row the gravitational field, and the bottom row the potential. Each column shows the rate using a different error norm; from left to right, we show the $L_1$, $L_2$, and $L_{\infty}$ norms, respectively. The expansion factor at which the first caustic forms is indicated with a dashed vertical line.}
\label{singular_convergence}
\end{figure*}

The resulting convergence rates are displayed as a function of time in Figure \ref{singular_convergence}. Formally, our adopted PIC scheme should be second-order accurate in space, and it indeed achieves this up until first shell crossing, at least for the gravitational field and the potential. After $a_{\text{caustic}}$, however, the convergence rates for $\rho_{\bi{i}}$ and $\bi{g}_{\bi{i}}$ drop dramatically, in all of the norms we consider. The solution for $\rho_{\bi{i}}$ does not converge at all in the max norm, while the solution for $\bi{g}_{\bi{i}}$ converges only slowly.

In a sense, it is not surprising that we see poor convergence in these metrics after $a_{\text{caustic}}$. With perfectly cold initial conditions, there is nothing to limit the peak densities in the dark matter caustics that form after $a_{\text{caustic}}$ other than the finite numerical resolution employed. Indeed, we see in Figure \ref{singular_convergence} that the $L_{\infty}$-norm of $\rho$ actually diverges with $q_{\rho} < 0$ after $a_{\text{caustic}}$. These results are consistent with the analytic theory of caustic structure (in the non-gravitating case) from \cite{shandarin_1989}, which states that, after pancake formation, there should be a $\rho(x) \propto x^{-1/2}$ singularity in the density, and therefore a $g(x) \propto x^{1/2}$ cusp in the gravitational field. While this poor convergence in 1D is limited to the caustic positions, it points towards more significant problems in 2D or 3D calculations, where the symmetry constraints of the initial conditions are not automatically met. 

\subsection{2D Results}
\label{singular_results_2D}

\begin{deluxetable}{ccc}
\tablecaption{\label{2D_singular_table} Summary of parameters for the singular, 2D runs}
\tablewidth{0pt}
\tablehead{
\colhead{$N_{\text{cells}}$} & \colhead{$n_x$} & \colhead{$C_{\text{exp}}$}
}
\startdata
    $128^2$ & ${1/2}^2$ & $2.0 \times 10^{-2}$ \\
    $128^2$ & $1^2$ & $2.0 \times 10^{-2}$ \\
    $128^2$ & $16^2$ & $2.0 \times 10^{-2}$ \\
    $256^2$ & ${1/2}^2$ & $1.0 \times 10^{-2}$ \\
    $256^2$ & $1^2$ & $1.0 \times 10^{-2}$ \\
    $256^2$ & $16^2$ & $1.0 \times 10^{-2}$ \\
\enddata
\end{deluxetable}

In this section, we examine a 2D analog of the non axis-aligned plane wave collapse problem considered by \cite{melott_demonstrating_1997}. This setup, in which the perturbation wave vector is not aligned with any of the coordinate axes, is known to produce unphysical clumping along directions in which the density should be constant given the symmetry of the initial conditions. In this section, we confirm that this effect is also present in 2D using our standard PIC method. We set the perturbation wavenumber $\bi{k} = (k_x, k_y)$ to be $(2, 5)$ in units of the fundamental mode. Otherwise, the initial conditions are the same as in our 1D resolution study: $a_{\text{ini}} = 1 / 200$, $a_{\text{caustic}} = 0.1$, and $a_{\text{stop}} = 1$. Throughout the rest of this paper, we refer to these initial conditions as the ``oblique" pancake.

We perform six runs in total, as summarized in Table \ref{2D_singular_table}. We vary the number of cells in the Poisson mesh over \{$128^2$, $256^2$\}, and run the problem with $n_x = $ $1/2^2$, $1$, and $16^2$ particles per Poisson cell. The particles are initially arranged within the Poisson cells in the usual way. In the run with $1/2^2$ particles per cell, we coarsen the Poisson mesh by a factor of two and place particles at centers of each cell in the resulting coarsened mesh. In the run with $16^2$ particles per cell, we refine the mesh, instead.  

\begin{figure*}
\epsscale{1.15}
\plotone{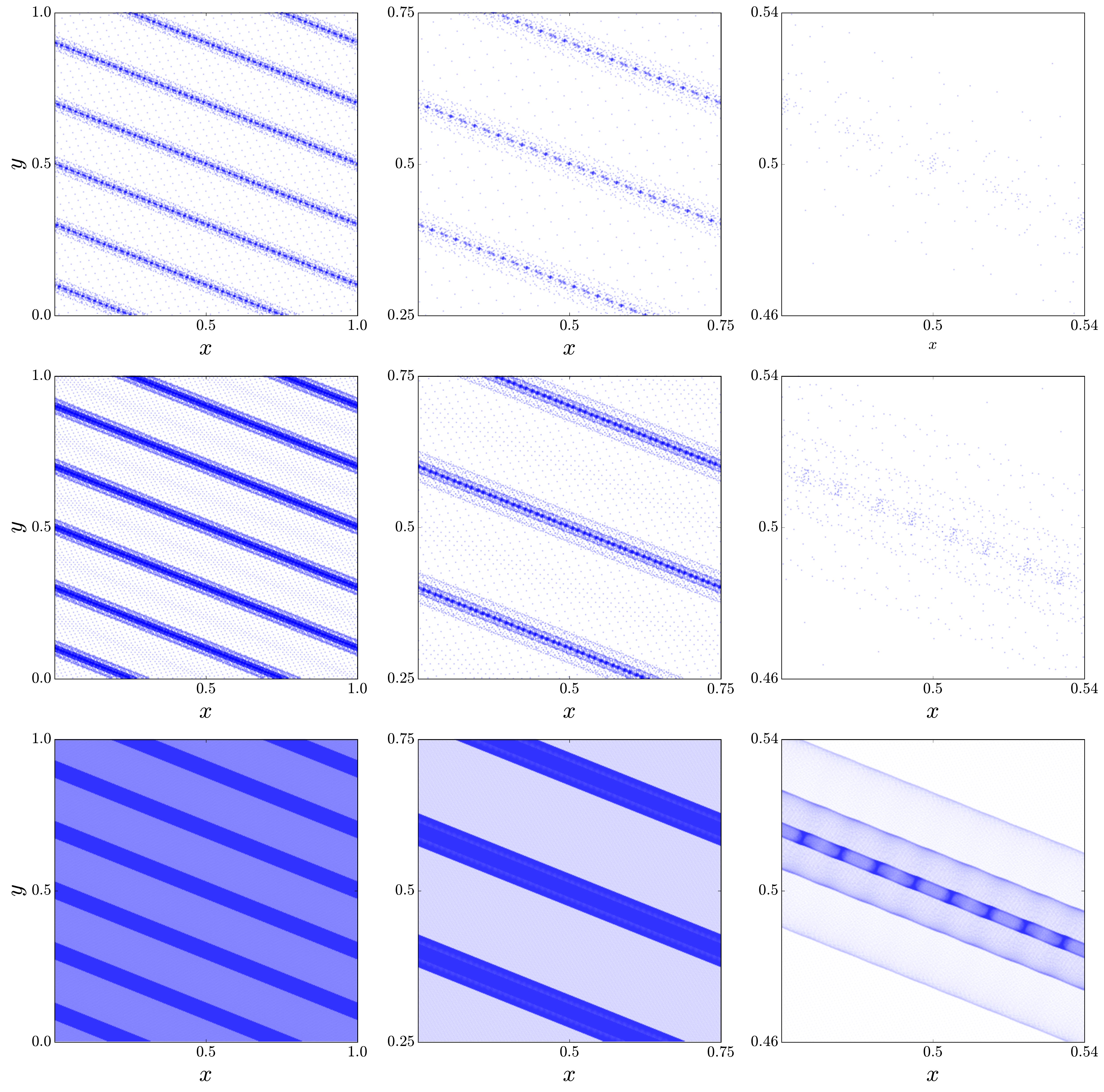}
\caption{Particle $x$ and $y$ positions at $a(t) = 1.0$ on the singular version of the oblique pancake problem. All runs use a $256^2$ Poisson mesh. The top row shows a run with a sparse particle count of 1 particle every 4 Poisson cells. The middle row uses 1 particle per cell, and the bottom uses a very high particle count of 256 particles per cell. For all three runs, the leftmost panels show the entire problem domain, and the view zooms in as you move to the right. All runs show signs of small-scale fragmentation that is inconsistent with the symmetry of the problem.}
\label{particle_plot_singular}
\end{figure*}

The resulting particle $x$ and $y$ positions at $a = 1$ are shown in Figure \ref{particle_plot_singular}. We only display the high resolution results in Figure \ref{particle_plot_singular}; the low resolution results are qualitatively similar. By the symmetry of the problem, the particle density should be constant along directions perpendicular to the axis of the perturbation. However, all of the runs show signs of small-scale fragmentation that do not obey this symmetry requirement, and thus cannot be valid solutions to Vlasov-Poisson for the given initial conditions. This effect is not limited to runs with sparse particle counts, and is even present in the run with the $16^2$ particles per Poisson cell - a much larger number than typically used in realistic dark matter simulations. We conclude that simply using very large particle counts with the basic PIC scheme is not sufficient to prevent these errors.  

\begin{figure*}
\epsscale{1.2}
\plotone{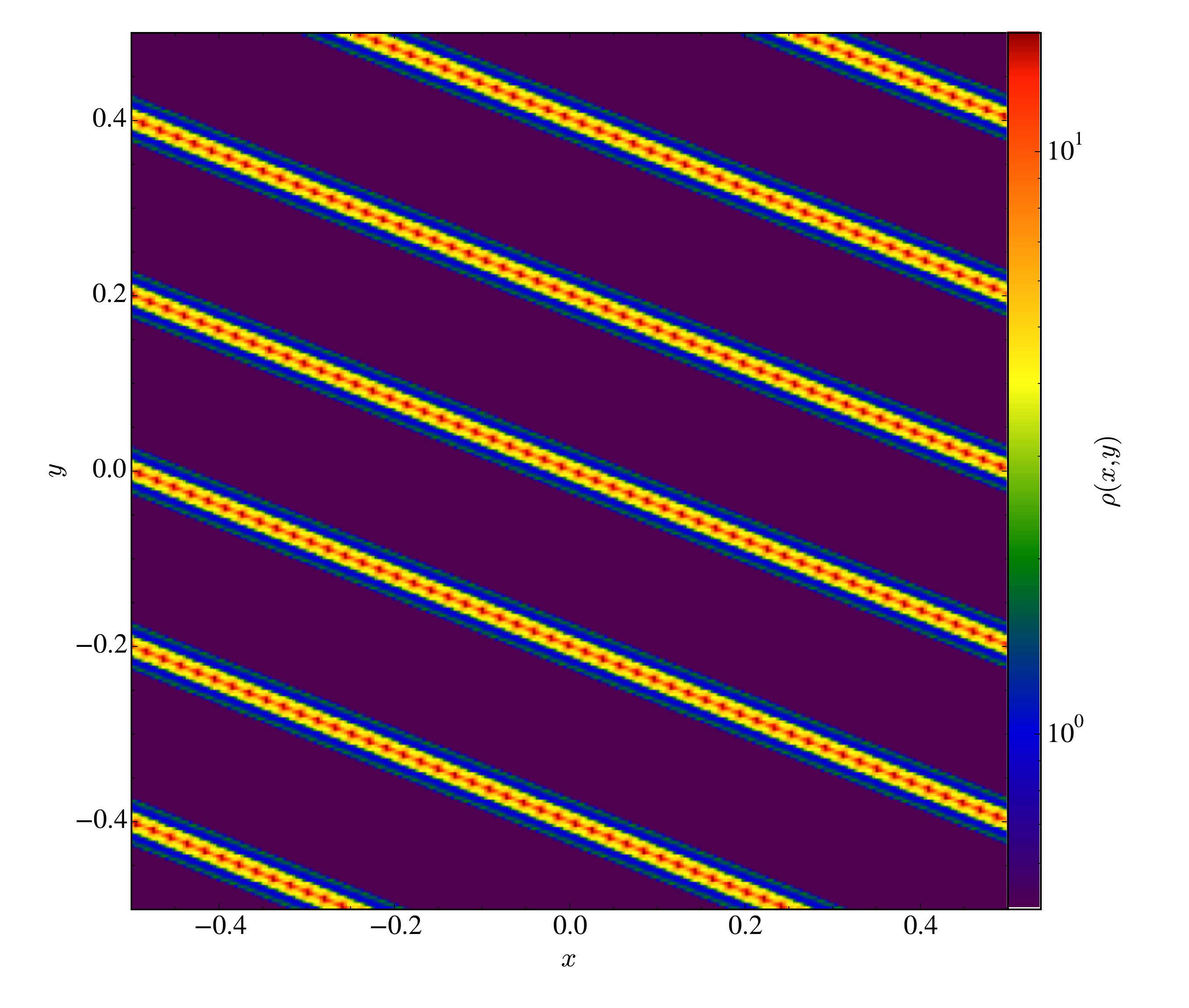}
\caption{The density field at $a(t) = 1$ from the low resolution, singular version of the oblique pancake problem. The density has been computed by deposition onto a $256^2$ mesh. }
\label{singular}
\end{figure*}

\begin{figure*}
\epsscale{1.2}
\plotone{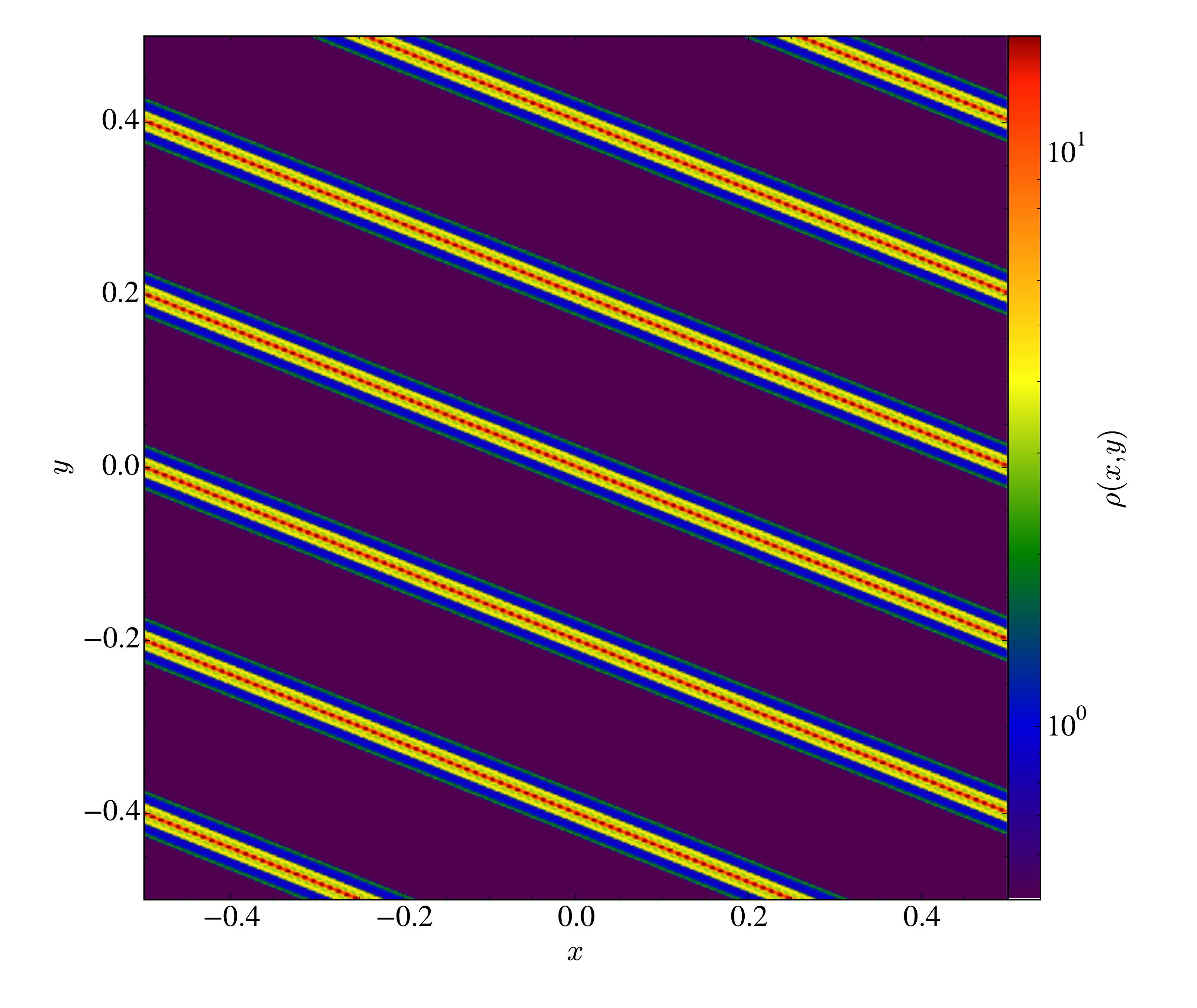}
\caption{The density field at $a(t) = 1$ from the high resolution, singular version of the oblique pancake problem, using the same color scale as Figure \ref{singular}. The density has been computed by deposition onto a $512^2$ mesh. }
\label{singular_hires}
\end{figure*}

This clumping effect is present in the grid-deposited density field, as well. In Figure \ref{singular}, we construct a 2$D$ density field by depositing the particles onto a $256^2$ mesh using cloud-in-cell deposition (Equation \ref{eq:particle_deposit}). This Figure was constructed using the run with $16^2$ particles per cell; the other runs produce an even stronger effect. Figure \ref{singular_hires} shows the corresponding density field from the $16^2$ particle per cell, high resolution run, where we generate the density by deposition onto a $512^2$ mesh. Both figures shows signs of unphysical fragmentation along the dark matter caustics. 

It is possible that this artificial fragmentation is related to the divergence of the density at the caustic locations in 1D observed in Section \ref{sec:singular_results_1D}. To test this hypothesis, we introduce a regularization procedure that limits the maximum density at the caustic positions. The results of this test are described in the next section.

\section{The Regularized Pancake}
\label{sec:regularized}

\begin{figure*}
\plotone{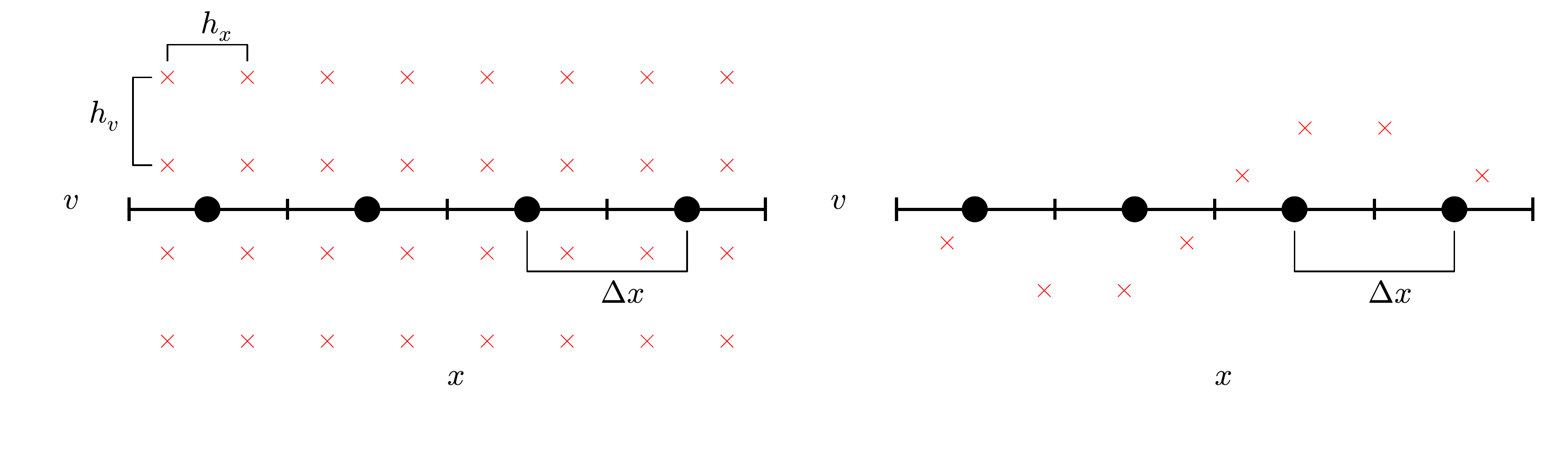}
\caption{A schematic demonstration of the difference between regularized and singular initial conditions. Left panel - the initial phase-space discretization for our regularized runs, for $\Delta x / h_x = 2$. The red crosses show the initial particle locations, and the black dots mark the centers of the Poisson mesh. The particle masses are computed by sampling the initial distribution function. Right - the initial particle positions for our singular runs, also for $\Delta x / h_x = 2$. The particles all have equal masses, and have been displaced from their initial positions using the Zel'dovich approximation.}
\label{discretization}
\end{figure*}

\begin{figure}
\epsscale{1.2}
\plotone{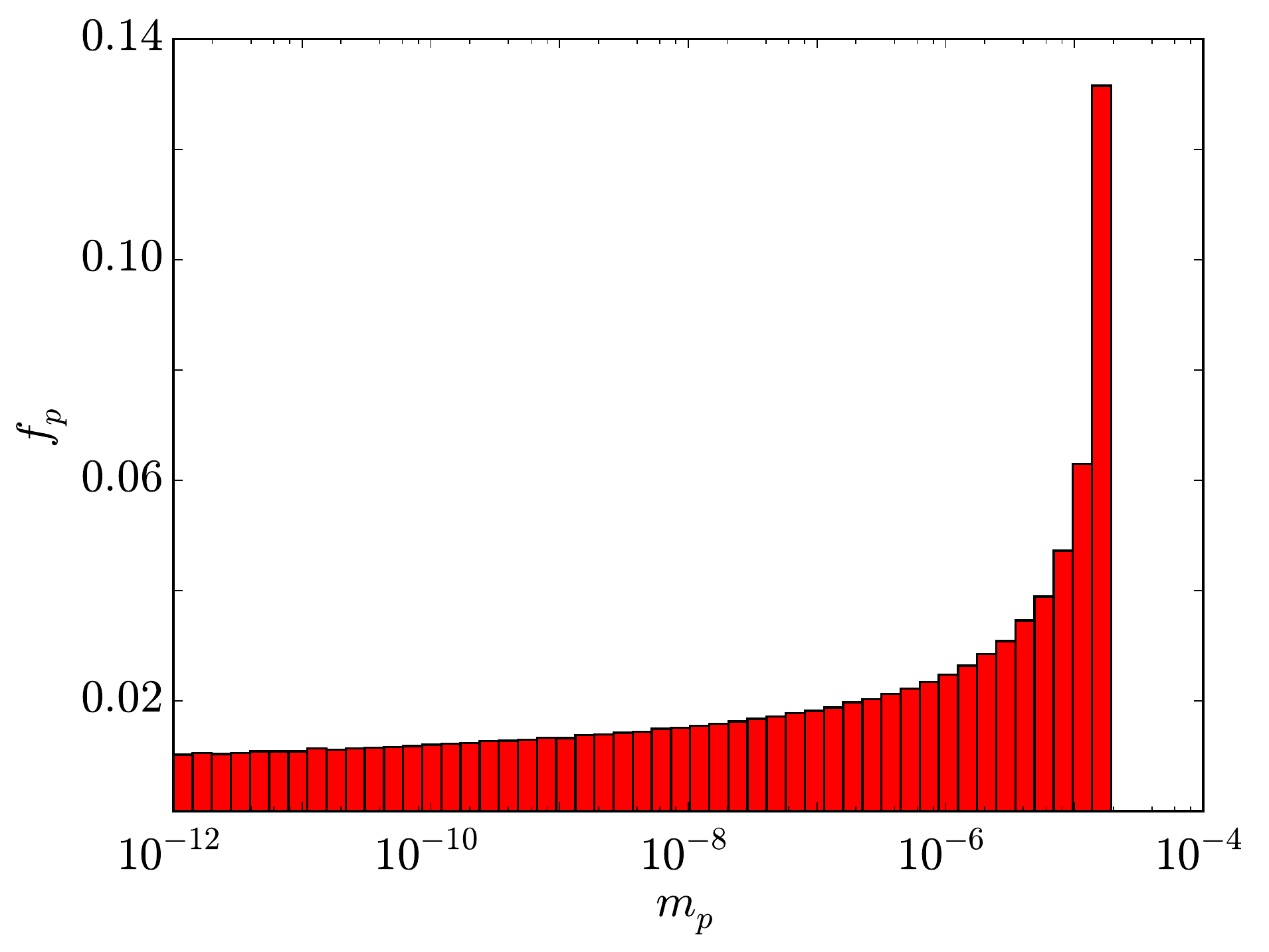}
\caption{A histogram of the particle masses obtained through the sampling procedure described in Section \ref{warm_ics}. We have used 50 logarithmically-spaced mass bins, and show $f_p$, the fraction of the total number of particles in each bin. To generate this plot, we have used $\sigma_i = 1.0$, $V = 6.0$, and taken $N_x = N_v = 512$. All other parameters are the same as in Section \ref{sec:singular_results_1D}.}
\label{mass_spectrum}
\end{figure}

\subsection{Initial Conditions}
\label{warm_ics}

The above initial conditions are singular in the sense that the initial
distribution function is a delta function in velocity space, centered at $\bi{v} = \bi{v}_z$. With delta function initial conditions, the solution to the Zel'dovich pancake problem at late times contains singularities in the density, and corresponding discontinuities in the derivative of the gravitational field, at the positions of the dark matter caustics. When solving this problem numerically with PIC or other particle methods, however, the peak densities obtained in the numerical solution will be determined by the finite amount of resolution employed. In effect, the mass deposition kernel, which spreads out the mass in a given particle over a few grid cells, provides a regularization that removes these singularities; however, it does so in a relatively uncontrolled way that is explicitly resolution-dependent. As a consequence, with perfectly cold initial data, it is not possible to separate numerical errors, which should improve with mesh refinement, from parameters of the numerical model, which should have well-defined effects on the solution once a converged answer is found.

Instead of letting the mesh spacing provide the regularization, we investigate an alternative regularization procedure that removes the singularities in the solution by introducing a finite initial velocity dispersion, $\sigma_i$, to equation \ref{cold_f}. We choose a Gaussian form for the velocity profile, so that 
the initial distribution becomes:
\begin{align}
\label{eq:regularized_Zel'dovich}
f(\bi{x}, \bi{v}, t_{\text{ini}}) &= \nonumber \\
& \left( \frac{1}{2 \pi {\sigma_i}^2} \right)^{D/2} \exp \left[-\frac{(\bi{v} - \bi{v}_z(\bi{x}, t_{\text{ini}}))^2}{2 {\sigma_i}^2} \right] \nonumber \\
&\times \frac{\langle \rho \rangle}{1 + a A \lvert \bi{k} \rvert \cos(\bi{k} \cdot \bi{q})}.
\end{align} 

This process makes the initial conditions more closely resemble those used in electrostatic PIC calculations, for which the expected convergence rates can be demonstrated. In the limit $\sigma_i \rightarrow 0$, Equation \ref{cold_f} is recovered. Using these modified initial conditions, we then evolve the regularized version of the Zel'dovich Pancake problem using the same PIC method as before. To generate the initial particles, we sample Equation \ref{eq:regularized_Zel'dovich} with a set of particles that are initially laid out on a cell-centered, Cartesian grid in phase space. We label this grid $\Omega_0$. The computational domain stretches from $0$ to $1$ in each physical dimension and $-V$ to $V$ in each velocity dimension. The domain bounds in velocity space are chosen so that $f(\bi{x}, \bi{v})$ is negligibly small outside of the problem domain. The number of points in $\Omega_0$ in the position and velocity dimensions are $N_x$ and $N_v$, respectively.  The locations of the cell centers are thus $\bi{x}_{\bi{i}} = (\bi{i} + 1/2) h_x$ and $\bi{v}_{\bi{j}} = (\bi{j} + 1/2) h_v - V$, where $(\bi{i}, \bi{j}) \in (\mathbb{Z}^D, \mathbb{Z}^D)$, and $(h_x, h_v) = (1 / N_x, 2V / N_v)$ are the cell spacings in position and velocity space. We place one particle in each cell $(\bi{i}, \bi{j})$. For a given initial distribution function, $f(\bi{x}, \bi{v}, t_{\text{ini}})$, the initial discretization is completed by assigning each particle $p \in \mathbb{P}$ a mass $m_p$:

\begin{equation}
m_p = f(\bi{x}_p^i, \bi{v}_p^i, t_{\text{ini}}) h_x^D h_v^D.
\end{equation} where $\bi{x}_p^i$ and $\bi{v}_p^i$ are the initial position and velocity of particle $p$. For computational efficiency, we discard particles with masses less than $10^{-12}$; experimentation reveals that tightening this value does not meaningfully affect our convergence results. We illustrate the difference between our ``regularized" and ``singular" initial particle layouts in Figure \ref{discretization}. We also show, in Figure \ref{mass_spectrum}, a sample spectrum of the particle masses obtained through this procedure. Specifically, these were the actual particle masses used in the $\sigma_i = 1$, $N_{\text{cells}} = 256$ calculation described in Table \ref{1D_table_reg}. Note that there is long tail of low-mass particles that contribute little to the overall distribution function - this is a consequence of our selecting a relatively low particle mass threshold and a relatively large domain in velocity space. It is likely that, by relaxing these parameters, significantly fewer particles can be used without degrading the solution.

As an aside, although our regularized initial conditions are ``warm'' in a sense, they should not be confused with ``Warm Dark Matter''. Our initial velocity dispersion is artificially large and introduced for numerical purposes only. It has nothing to do with the physical velocity
dispersion of a hypothetical DM candidate. Simulations of structure
formation in Warm DM universes in fact use ``singular'' initial conditions
by our definition, but they suppress power in fluctuations below some
free-streaming scale associated with the rest mass of the assumed dark
matter particle.

\subsection{1D Results}
\label{sec:warm_results_1D}

\begin{deluxetable}{ccccc}[!H]
\tablecaption{\label{1D_table_reg} Summary of parameters for the regularized, 1D pancake runs}
\tablewidth{0pt}
\tablehead{
\colhead{$\sigma_i$ \tablenotemark{a}} & \colhead{$N_{\text{cells}}$} & \colhead{$N_x$ \tablenotemark{b}} & \colhead{$N_v$ \tablenotemark{c}} & \colhead{$C_{\text{exp}}$}
}
\startdata
    $\{1, 1/2, ..., 1/32\}$ & $64$ & $128$ & 128 & $4.0 \times 10^{-2}$ \\
    $\{1, 1/2, ..., 1/32\}$ & $128$ & $256$ & 256 & $2.0 \times 10^{-2}$ \\
    $\{1, 1/2, ..., 1/32\}$ & $256$ & $512$ & 512 & $1.0 \times 10^{-2}$ \\
    $\{1, 1/2, ..., 1/32\}$ & $512$ & $1024$ & 1024 & $5.0 \times 10^{-3}$ \\
    $\{1, 1/2, ..., 1/32\}$ & $1024$ & $2048$ & 2048 & $2.5 \times 10^{-3}$ \\
    $\{1, 1/2, ..., 1/32\}$ & $2048$ & $4096$ & 4096  & $1.25 \times 10^{-2}$ \\
\enddata
\tablenotetext{a}{The notation $\{1, 1/2, ..., 1/32\}$ means we have varied the parameter $\sigma_i$ over the stated range.}
\tablenotetext{b}{The number of cells in the initial particle mesh in the $x$ direction.}
\tablenotetext{c}{The number of cells in the initial particle mesh in the $v$ direction.}
\end{deluxetable}

\begin{figure}
\epsscale{1.2}
\plotone{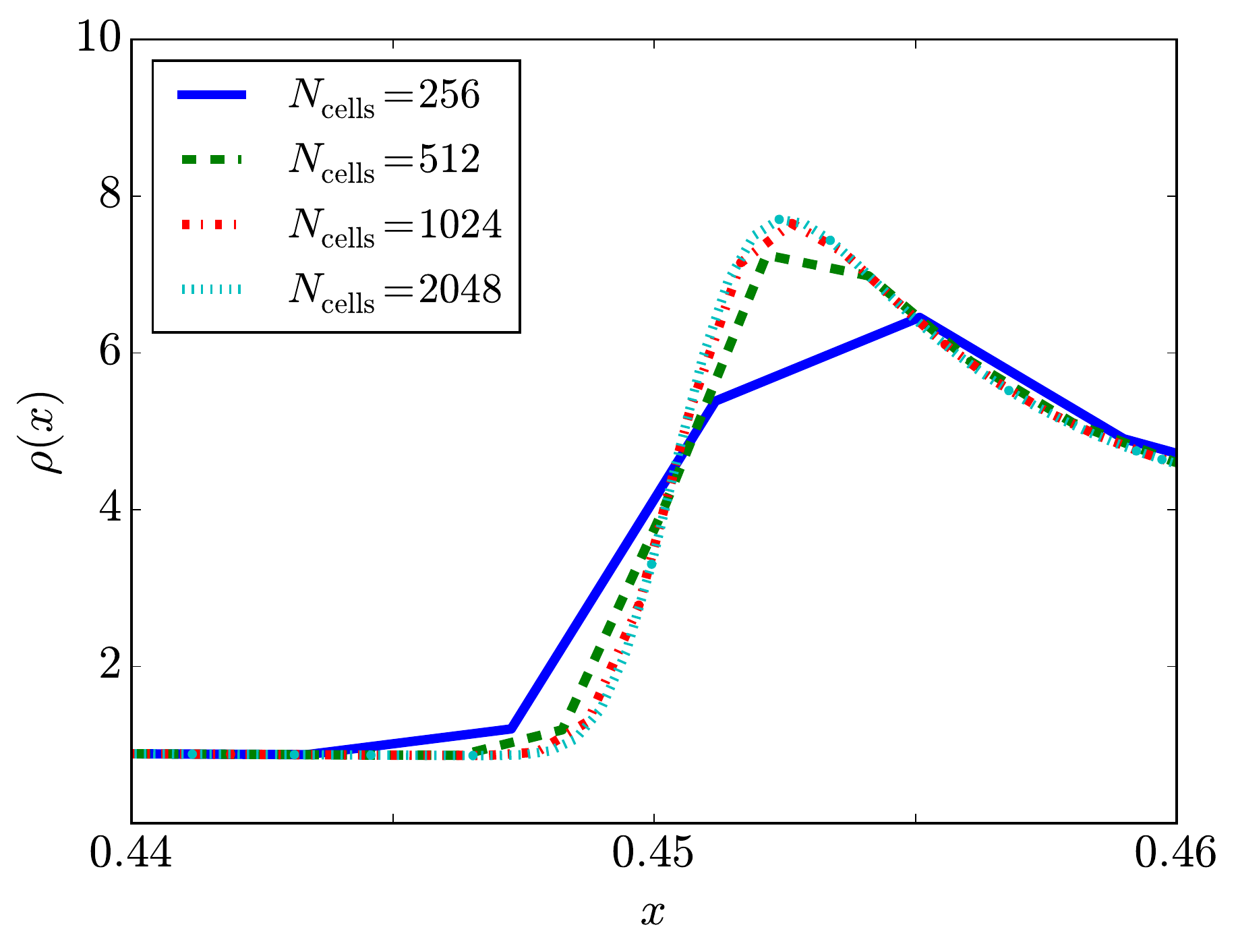}
\caption{A zoomed-in view of one of the outer density caustics from the 1D, regularized problem with $\sigma_i = 1$, taken at $a(t) = 1$. The different colors correspond to different resolutions: solid blue line - $N_{\text{cells}} = 256$; dashed green line - $N_{\text{cells}} = 512$; dashed-dotted red line - $N_{\text{cells}} = 1024$; dotted cyan line - $N_{\text{cells}} = 2048$. }
\label{1D_density_converge_regularized}
\end{figure}

\begin{figure*}
\epsscale{1.2}
\plotone{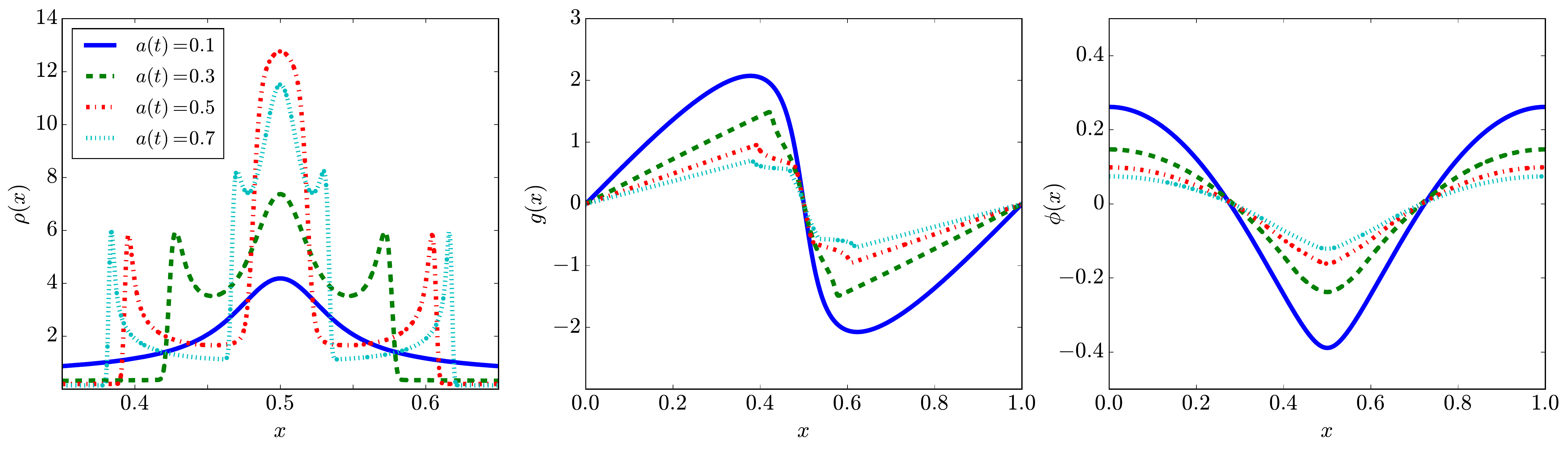}
\caption{The time evolution of the solution to the $\sigma_i = 1$ version of the 1D, regularized problem. The left panel shows the density, the middle panel the gravitational field, and the right panel the potential. The plotted curves are from the run with $N_{\text{cells}} = 2048$; the $N_{\text{cells}} = 1024$ run would be indistinguishable from the plotted solutions at the scales shown. The different lines correspond to the results at different expansion factors: solid blue line - $a(t) = 0.1$; dashed green line - $a(t) = 0.3$; dashed-dotted red line - $a(t) = 0.5$; dotted cyan line - $a(t) = 0.7$.}
\label{evolution_regularized}
\end{figure*}

To begin, we fix the regularization parameter at $\sigma_i = 1$. We otherwise repeat the calculation in Section \ref{singular}, using the same values for $a_{\text{ini}}$, $a_{\text{caustic}}$, $a_{\text{stop}}$, $k$, and $C_{\text{exp}}$. For our lowest resolution run, we use $N_{\text{cells}} = 256$, $N_x = 512 = N_v = 512$. These choices were motivated by the convergence theory of \citetalias{wang_particle--cell_2011}, which states that, for convergence, $\Delta x <= h_x$ and $h_x = O(h_v)$ (in our dimensionless formulation of the Vlasov-Poisson problem). While this doesn't require that $N_x = N_v$, we find it convenient to enforce this condition. This amounts to employing a relatively fine base grid, leading us to over-resolve the problem in velocity space, but this is not much of an issue in this 1+1D problem. We also perform runs with the resolution increased to $N_{\text{cells}} = 512$, $1024$, $2048$, and $4096$, and the other elements of the discretization refined accordingly. That is, when we refine the Poisson mesh, we also refine $N_x$, $N_v$, and decrease the time step parameter $C_{\text{exp}}$, as before. These run parameters are summarized in Table \ref{1D_table_reg}. The resulting solutions are shown in Figure \ref{evolution_regularized}. The effect of the (in this case quite large) artificial velocity dispersion to smooth out the density peaks over some comoving length scale, so that the peak densities in the caustics now have well-defined maximum values. As shown in Figure \ref{1D_density_converge_regularized}, the density caustics now appear to be converging with increased mesh resolution. 

\begin{figure*}
\epsscale{1.15}
\plotone{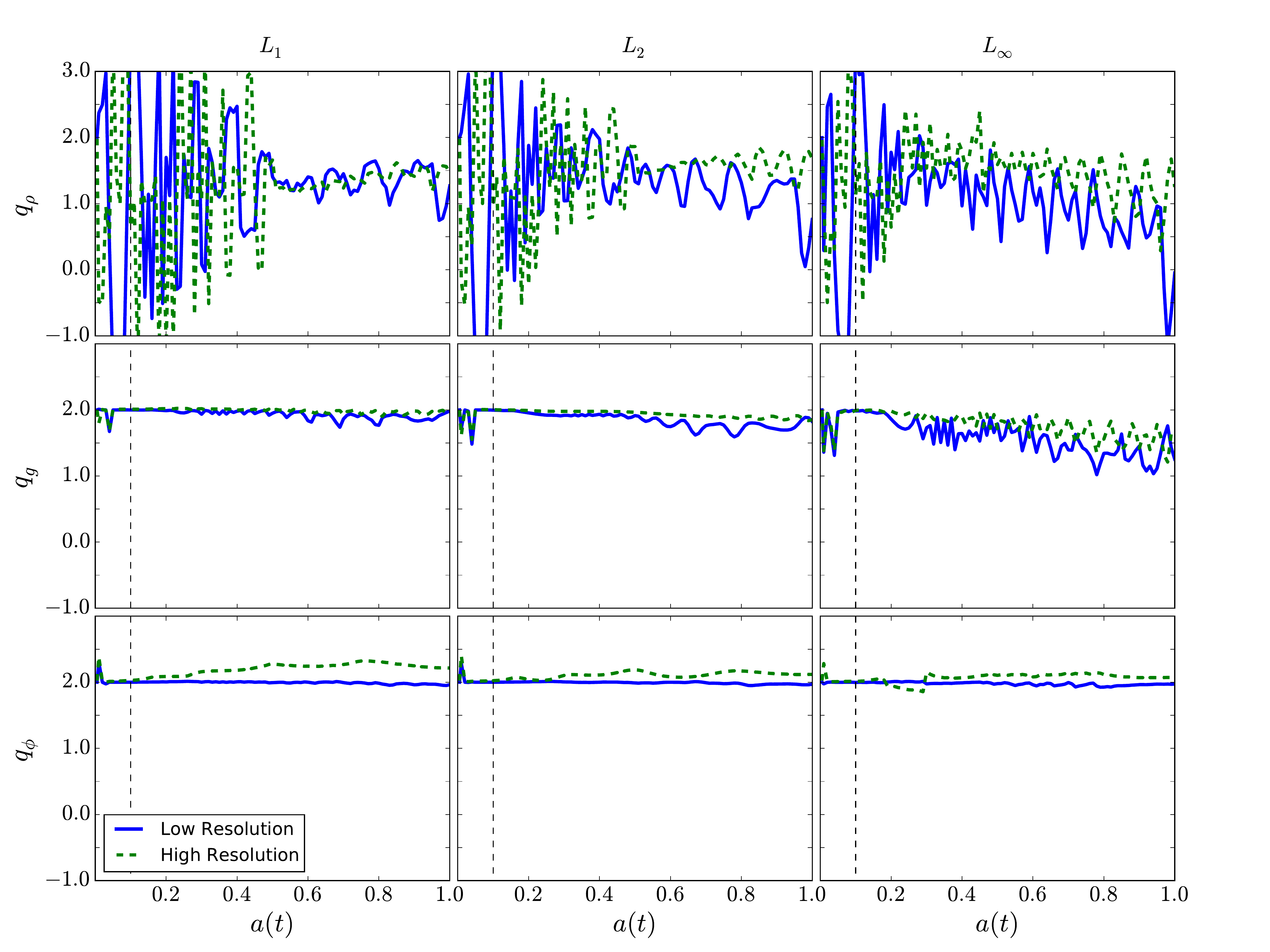}
\caption{The same as Figure \ref{singular_convergence}, but for the $\sigma_i = 1$ version of the regularized problem. The blue line compares the result for the $N_{\text{cells}} = 256, 512,$ and $1024$, while the green line shows the same quantity for resolutions $N_{\text{cells}} = 512, 1024,$ and $2048$.}
\label{regularized_convergence}
\end{figure*}

We also examine the convergence rates for the density, gravitational field, and potential on the regularized version of the problem. The rates are computed as in Section \ref{sec:singular_results_1D} and shown in Figure \ref{regularized_convergence}. Overall, the convergence behavior is much better than on the singular version of the problem. The density order is quite noisy at early times, but the rates for the potential and gravitational field are close to 2 for all the times we consider. The convergence rate for the density is slower than 2 at late times, but the density is not divergent as it was on the singular problem.

\subsection{The Double Limit}
\label{sec:double_limit}
\begin{figure*}
\epsscale{1.15}
\plotone{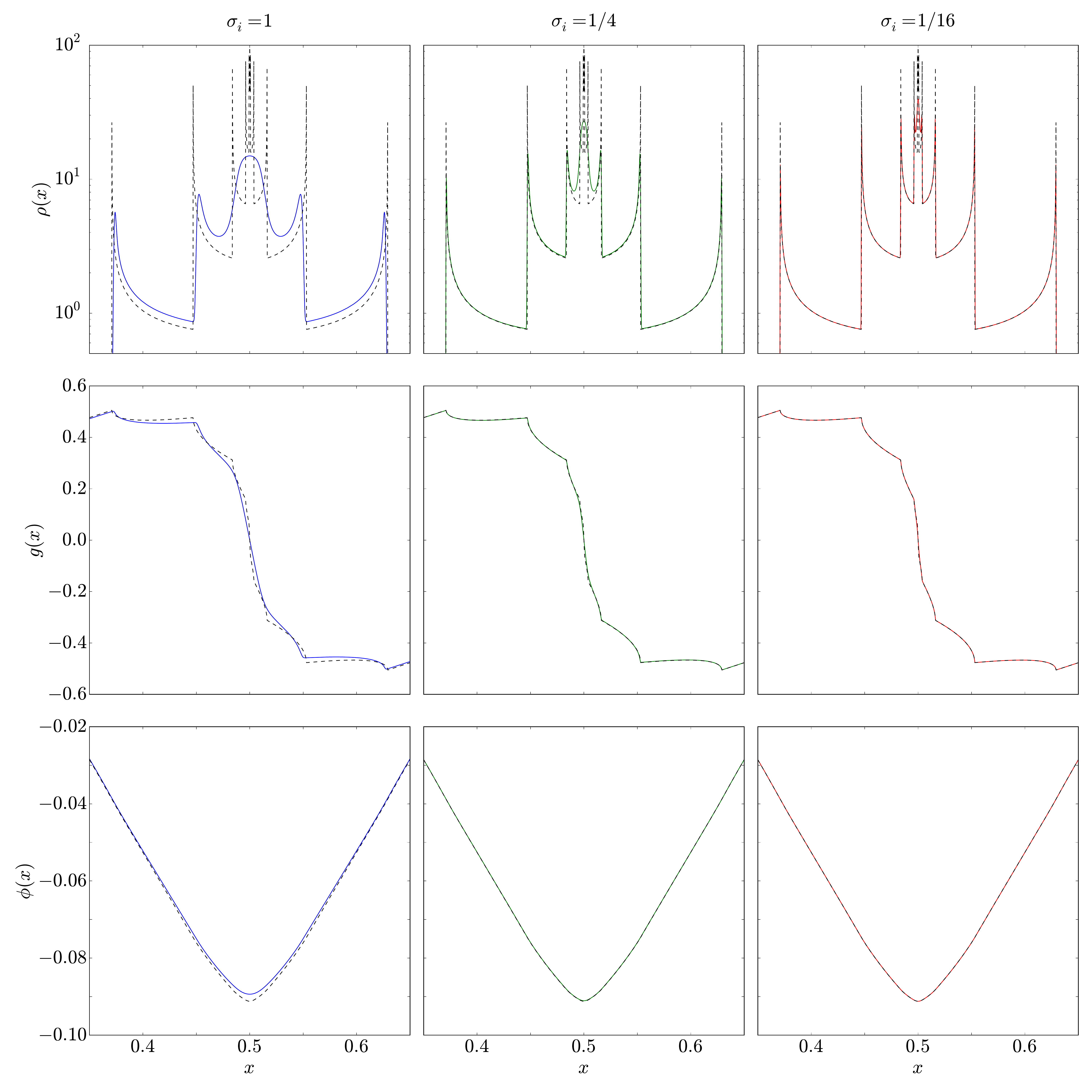}
\caption{The effect of letting $\sigma_i \rightarrow 0$ on the obtained solutions. The solid lines show the converged density (top row), gravitational field (middle row), and potential (bottom row) from our regularized runs at $a(t) = 1$. The black dotted lines are taken from a very high ($N_{\text{cells}} = 16,384$) resolution singular calculation. The regularized results are taken from the runs with $N_{\text{cells}} = 4096$ - the highest resolution available. The solutions are sufficiently converged that the $N_{\text{cells}} = 2048$ solutions would be indistinguishable from the plotted curves on this plot. The left column (blue) shows $\sigma_i = 1$, the middle column (green) shows $\sigma_i = 1/4$, and the right column (red) shows $\sigma = 1/16$.}
\label{regularization_limit}
\end{figure*}

In Section \ref{sec:warm_results_1D}, we showed that for a given value of $\sigma_i$, our PIC method converges at late times on the 1D, regularized pancake problem. Of course, in dark matter simulations, we are really interested in the case where $\sigma_i = 0$. Fortunately, we can exploit the fact that we are able to obtain converged solutions to the regularized problems to also obtain a solution to the singular problem by considering the limit as $\sigma_i \rightarrow 0$. This procedure gives us a way of separating numerical errors, which should improve with mesh refinement if our method converges, from regularization error, i.e. the fact that our caustics are artificially smoothed out by the artificial initial velocity dispersion. In this section, we illustrate this process on our 1D problem setup. We vary the quantity $\sigma_i$ over the range \{$1$, $1/2$, $1/4$, $1/8$, $1/16$, $1/32$\}. For each value of $\sigma_i$, we increase the mesh resolution until we obtain a converged solution (see Table \ref{1D_table_reg} for a full list of the run parameters). Next, we consider the limit of the converged solutions to the regularized problems as $\sigma_i \rightarrow 0$. This concept of a ``double limit" - letting both the mesh spacing and $\sigma_i$ go to zero - was inspired by \cite{krasny_desingularization_1986}, who used a similar procedure to remove singularities in vortex methods for incompressible fluid flows. The results are shown in Figure \ref{regularization_limit}. These figures show that the regularized solutions approach the singular solution as $\sigma_i \rightarrow 0$. The main difference between the singular and regularized runs is that the peak caustic densities are limited in the regularized runs, and the structures in the inner caustics are smoothed out over some comoving length scale. 

In particular, as $\sigma_i$ is decreased, we resolve more and more of the internal structure of the dark matter density distribution. To make this more quantitative, we associate a comoving length scale $\bi{d}$ with each solution as follows. For each value of $\sigma_i$, we compute the comoving distance (at $a(t) = 1$) between the positions of the outermost caustics that are present in the perfectly cold run, but not present in the corresponding regularized run. To identity the peak positions, we use a standard clump finding algorithm. The results are as follows. For the largest value of $\sigma_i = 1$, even the outermost caustic positions differ slightly from the results obtained in the perfectly cold limit, while the internal caustics have been completely washed out, such that none of the caustics with positions within $\bi{d} \approx 0.032$ of the center in the cold problem are present in the warm problem. For $\sigma_i = 1/4$, the outermost caustic positions match precisely, while caustics within $\bi{d} \approx 0.008$ of the center in the cold problem have smoothed out. For $\sigma_i = 1/16$, the corresponding length scale is $\bi{d} \approx 0.0013$.

Outside of the caustics, the effect of $\sigma_i$ on the solutions is small. This finding suggests an alternative approach to perfectly cold initial conditions in DM simulations. Instead of using singular initial conditions, which make code validation difficult, one instead decides on a length scale below which one will not believe the answer, and then chooses the corresponding initial artificial velocity dispersion. Given that velocity dispersion, one can rigorously demonstrate convergence at the desired order of accuracy.

\begin{figure*}
\epsscale{1.2}
\plotone{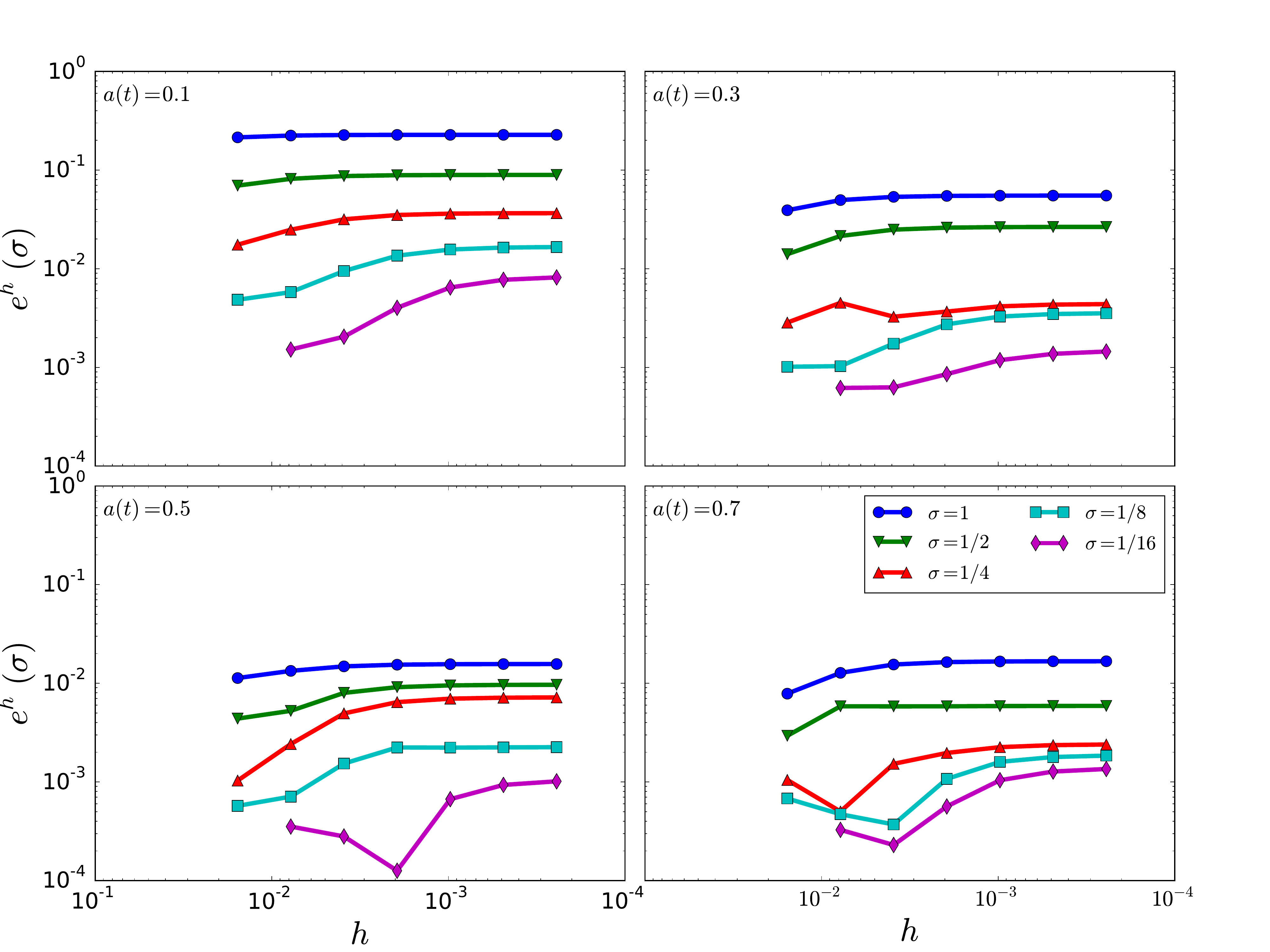}
\caption{Our error metric $e^h(\sigma)$ computed for all runs in Table \ref{1D_table_reg}. The blue curve shows  $e^h(\sigma = 1)$ - that is, the $L_2$ norm of the change in the solution for the gravitational field when $\sigma$ changes from 1.0 to 0.5, as a function of $h$. The green curve corresponds to $\sigma = 0.5$, the red curve $\sigma = 0.25$, cyan is $\sigma = 0.125$, and magenta is $0.0625$. The values in this plot were computed at $a(t) = 0.1$, (top left), $a(t) = 0.3$, (top right), $a(t) = 0.1$, (bottom left), and $a(t) = 0.1$, (bottom right).}
\label{double_limit}
\end{figure*}

To make the concept of the double limit more quantitative, we define the following error metric, analogous to the Richardson extrapolated error introduced in Section \ref{sec:singular_results_1D}:
\begin{equation}
e^h(\sigma) = \lvert \lvert \bi{g}_i^h(\sigma)  - \bi{g}_i^{h}(\sigma / 2) \rvert \rvert.
\end{equation} That is, for each value of the mesh spacing $h$, we compute the norm of the difference between the solution with the regularization parameter $\sigma_i = \sigma$ and the one with $\sigma_i  = \sigma / 2$. We use the $L_2$ norm to compute the error metric, and focus on the solution for the gravitational field. The results for $e^h(\sigma)$ for all the runs in Table \ref{1D_table_reg} are shown in Figure \ref{double_limit}, for the same expansion factors shown in Figure \ref{evolution_singular}. There are two conclusions to be drawn from this plot. First, it shows that the quantity $e^h(\sigma)$ approaches a well-defined value for each pair of velocity dispersions. By $N_{\text{cells}} = 2048$, $e^h(\sigma)$ has leveled off for all of the values of $\sigma_i$ considered. Second, it confirms that the regularized solutions themselves converge as $\sigma_i \rightarrow 0$. 

There are several conclusions that can be drawn from this exercise. First, it shows that our method of regularizing the initial conditions through an artificial Gaussian velocity dispersion does not affect the dark matter structure outside the caustic positions, provided $\sigma_i$ is chosen small enough. Second, it gives us reason to believe that the solution to the singular problem obtained via the standard PIC method is in fact the correct answer to the 1D pancake problem after $a_{\text{caustic}}$. Because the density field diverges in the perfectly cold limit, the standard method of validating the solution through examining the convergence rate is not available. By considering the limit of a series of regularized problems, we can have confidence that the solution obtained to the singular problem is correct. Third, since we are not able to solve the oblique version of this problem directly using cold initial conditions, regularization may provide a way to obtain approximate solutions to the singular problem in more than one spatial dimension. We test this hypothesis in the next section.

\subsection{2D Results}
\label{warm_results_2D}

\begin{deluxetable}{ccccc}
\tablecaption{\label{2D_regularized_table} Summary of parameters for the regularized, 2D pancake runs}
\tablewidth{0pt}
\tablehead{
\colhead{$\sigma_i$} & \colhead{$N_{\text{cells}}$} & \colhead{$N_x$ \tablenotemark{a}} & \colhead{$N_v$ \tablenotemark{b}} & \colhead{$C_{\text{exp}}$}
}
\startdata
     $1/16$ & $128^2$ & $256$ & 64 & $2.0 \times 10^{-2}$ \\
     $1/16$ & $256^2$ & $512$ & 128 & $1.0 \times 10^{-2}$ \\
\enddata
\end{deluxetable}

\begin{figure*}
\epsscale{1.15}
\plotone{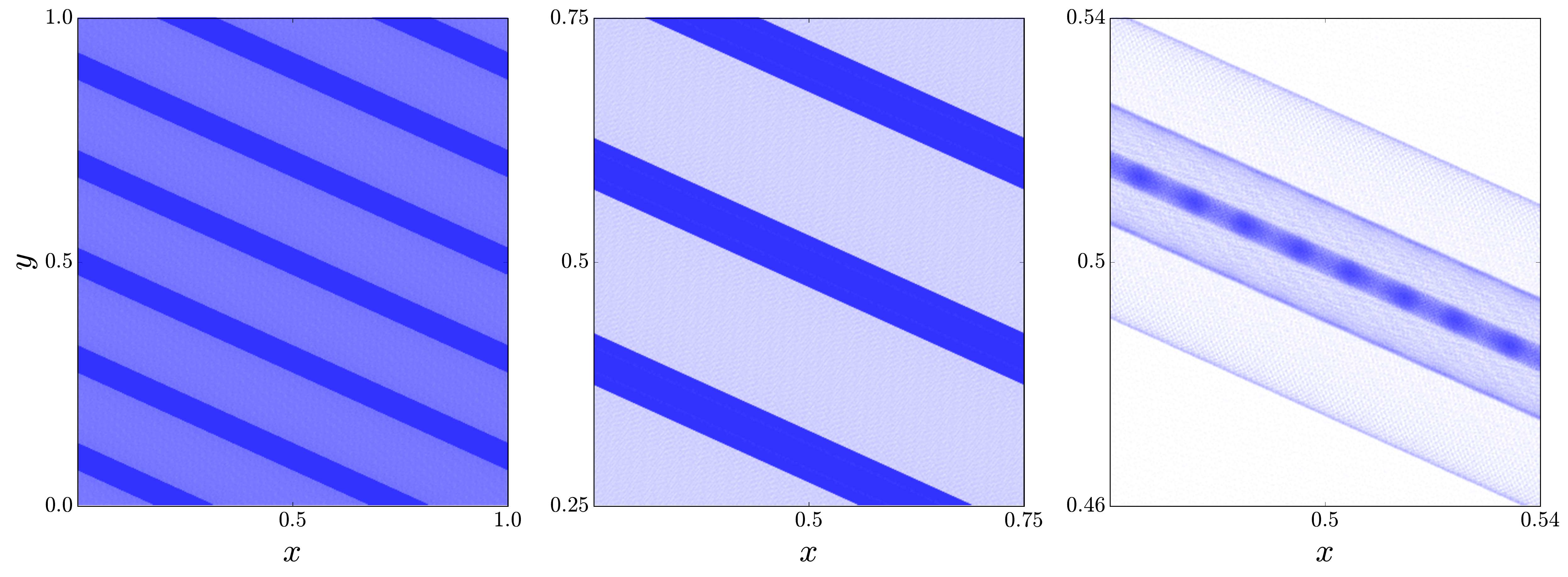}
\caption{Particle $x$ and $y$ positions at $a(t) = 1$ for the $\sigma_i = 1/16$ version of the regularized oblique pancake problem. The scales shown are the same as in Figure \ref{particle_plot_singular}}
\label{particle_plot_regularized}
\end{figure*}

\begin{figure*}
\epsscale{1.2}
\plotone{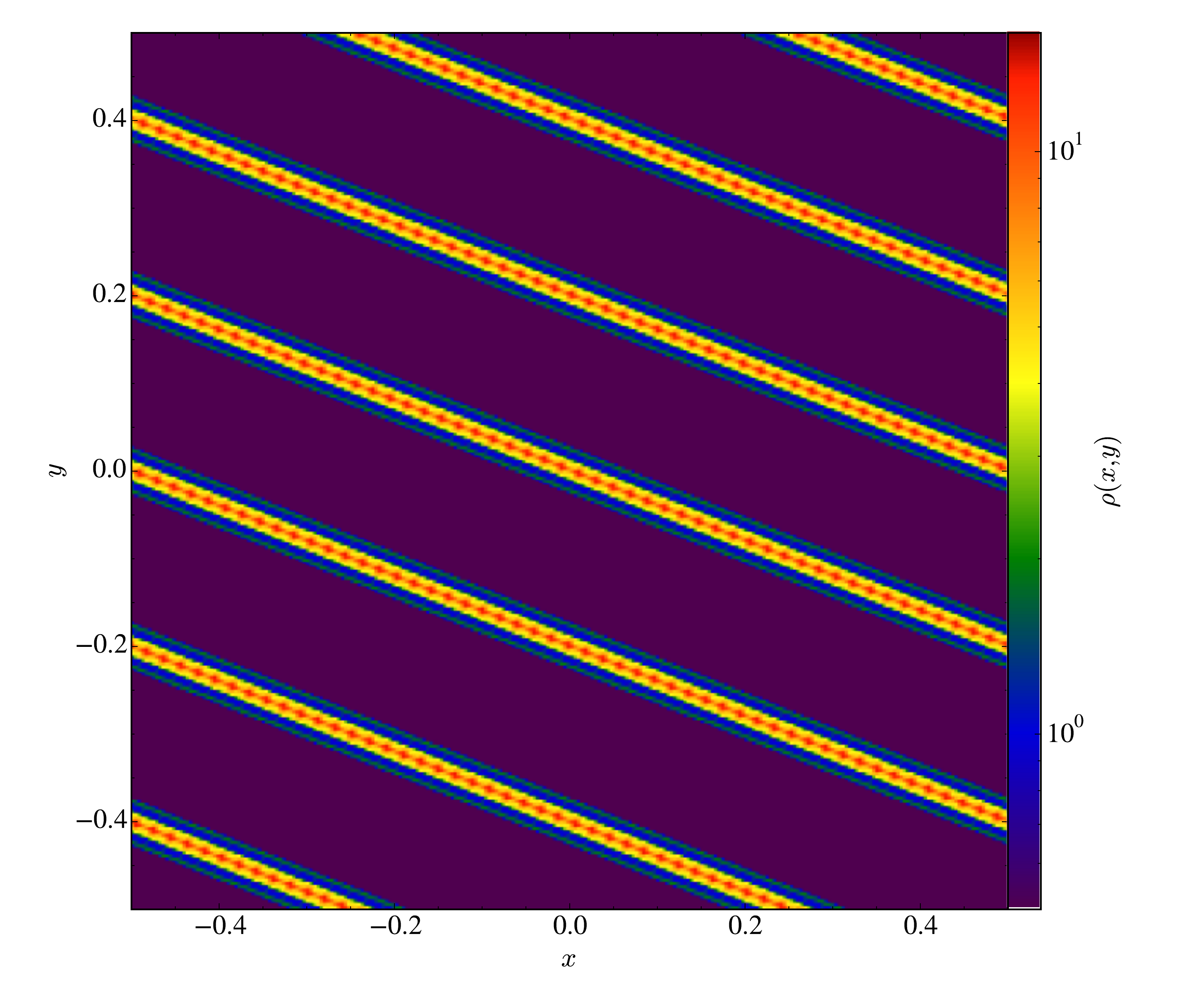}
\caption{The solution for the density at $a=1$ from the low-resolution calculation of the oblique pancake problem using regularized initial conditions with $\sigma_i = 1/16$. The color scale is the same as in Figure \ref{singular}. Particle remapping was not turned on for this run. For details of the parameter choices we adopted, see text}
\label{regularized}
\end{figure*}

\begin{figure*}
\epsscale{1.2}
\plotone{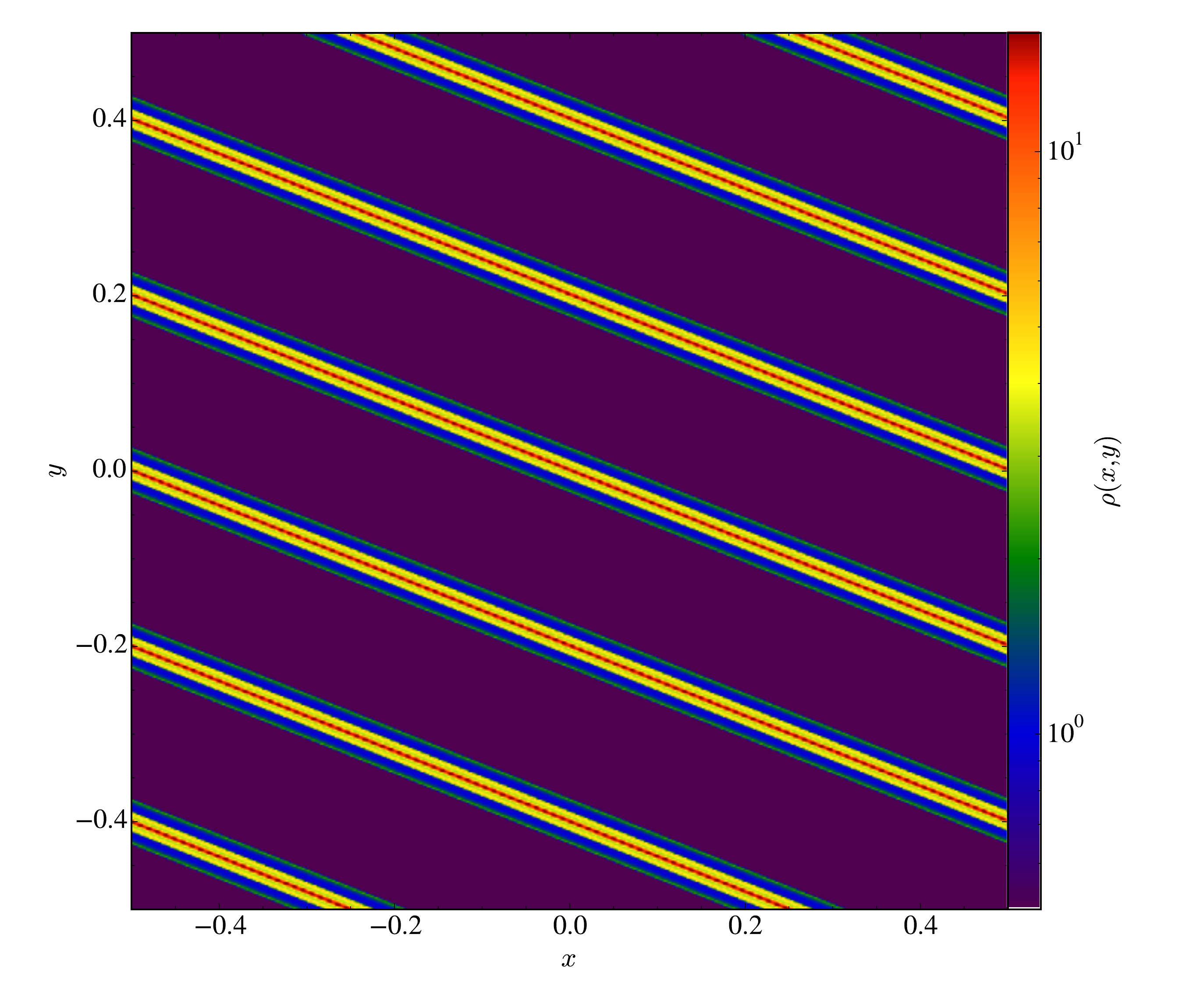}
\caption{The solution for the density at $a=1$ from the high-resolution calculation of the oblique pancake problem using regularized initial conditions with $\sigma_i = 1/16$. The color scale is the same as in Figure \ref{singular}. Particle remapping was not turned on for this run. For details of the parameter choices we adopted, see text.}
\label{regularized_hires}
\end{figure*}

We now compute the solution to a regularized version of the 2-dimensional problem in Section \ref{singular_results_2D}. We do two runs in this section, which are summarized in Table \ref{2D_regularized_table}. Both runs have set $\sigma_i = 1/16$, and construct $\Omega_0$ such that $N_x = 256$, and $N_v = 128$. In this case, the total number of cells in the phase space domain on which the particles are initialized is $N_x^2 \times N_v^2$. The number of cells in the Poisson mesh is fixed at $N_{\text{cells}} = 128^2$. All of the other parameters are the same as in Section \ref{singular_results_2D}. The resulting particle positions for the higher resolution are shown in Figure \ref{particle_plot_regularized}. We also display deposited fields, as in Section \ref{singular_results_2D}, for both the low (Figure \ref{regularized}) and high (Figure \ref{regularized_hires}) resolutions runs. 

Unfortunately, despite the improved convergence rates in 1D, the regularized problem suffers from the same clumping problem as the singular problem for the oblique pancake setup. However, there is an additional refinement we can introduce to the basic PIC procedure. In the next section, we introduce particle remapping, and show that, in concert with the regularization of the initial conditions described in this section, it can improve the performance of PIC on the oblique problem.  

\section{Remapping}    
\label{sec:remapping}
    
    In the error analysis of electrostatic PIC in \citetalias{wang_particle--cell_2011}, the stability error for the electric field contains a term that grows exponentially with time. Given long enough integration times, this error term can become large enough to significantly degrade the quality of the numerical solution. A common strategy for controlling this errors is to periodically
restart the problem with a new set of particles, before the accumulated
errors in the particle trajectories become too large. Such ``regridding'' or
``remapping'' techniques have been applied successfully to particle schemes for fluid dynamics, such as vortex
methods and smoothed particle hydrodynamics (SPH) \citep{vadlamani_particle-continuum_2004, koumoutsakos_inviscid_1997, cottet_vortex_2000, chaniotis_remeshed_2002}, and to PIC in the
context of plasma physics \citep{denavit_numerical_1972, chen_coarse-graining_2008, wang_particle--cell_2011, wang_adaptive_2012}. 

In the case of PIC, the basic idea is to
interpolate the distribution function back onto a regular mesh in phase
space, initialize a new set of particles, and restart the
calculation. In principle, the remapping mesh can uniform. However, repeatedly re-sampling the distribution 
function on a uniform mesh leads to a large number of relatively weak particles in the tails of the distribution function - a poor use of the available resolution elements. Furthermore, in the cosmological context, remapping serves an additional role. 
We have regularized the initial conditions by introducing a finite initial
velocity dispersion, \(\sigma_i\). However, as the scale factor $a(t)$
increases, $\sigma(t)$ will scale as $\sigma_i/a(t)$. Thus, if we
employ the above procedure an a uniform mesh, the artificial velocity spread in the
distribution function will be resolved more and more poorly as time
advances, undoing the natural adaptivity of Lagrangian schemes.

To remedy these issues, we remap on an adaptive hierarchy of meshes that automatically 
adjusts the velocity-space resolution on the finest level so that the distribution is resolved by approximately the same number of particles throughout the simulation. The AMR hierarchy is defined as a set of cell-centered, Cartesian meshes in phase space. We label the meshes $\Omega_{\ell}$, where $ 0 \leq \ell \leq \ell_{\text{max}} $ are the level numbers. The coarsest mesh, $\Omega_0$, covers the entire phase-space domain and is the same as the mesh used for generating the initial particles. The finer meshes are unions cell-centered rectangles and in general cover only a portion of the problem domain. The mesh spacing of level $\ell$ is related to the mesh spacing on the next coarsest level by $\bi{h}_{\ell} = \bi{h}_{\ell - 1} / r_{\text{ref}}$, where $r_{\text{ref}}$ is the AMR refinement ratio. The valid region of an AMR level is defined as the region not covered by finer levels, and the composite grid is defined as the union of the valid regions of each level.  

We describe the remapping algorithm in detail below, first using a single, uniform mesh to represent $f(\bi{x}, \bi{v})$, second using a fixed hierarchy of levels, and finally using the full, adaptive hierarchy. Note that, as presently constructed, the remapping procedure requires the regularization of the initial conditions - there must be a well-defined distribution function to generate the particles from. 

\begin{figure}
\epsscale{1.2}
\plotone{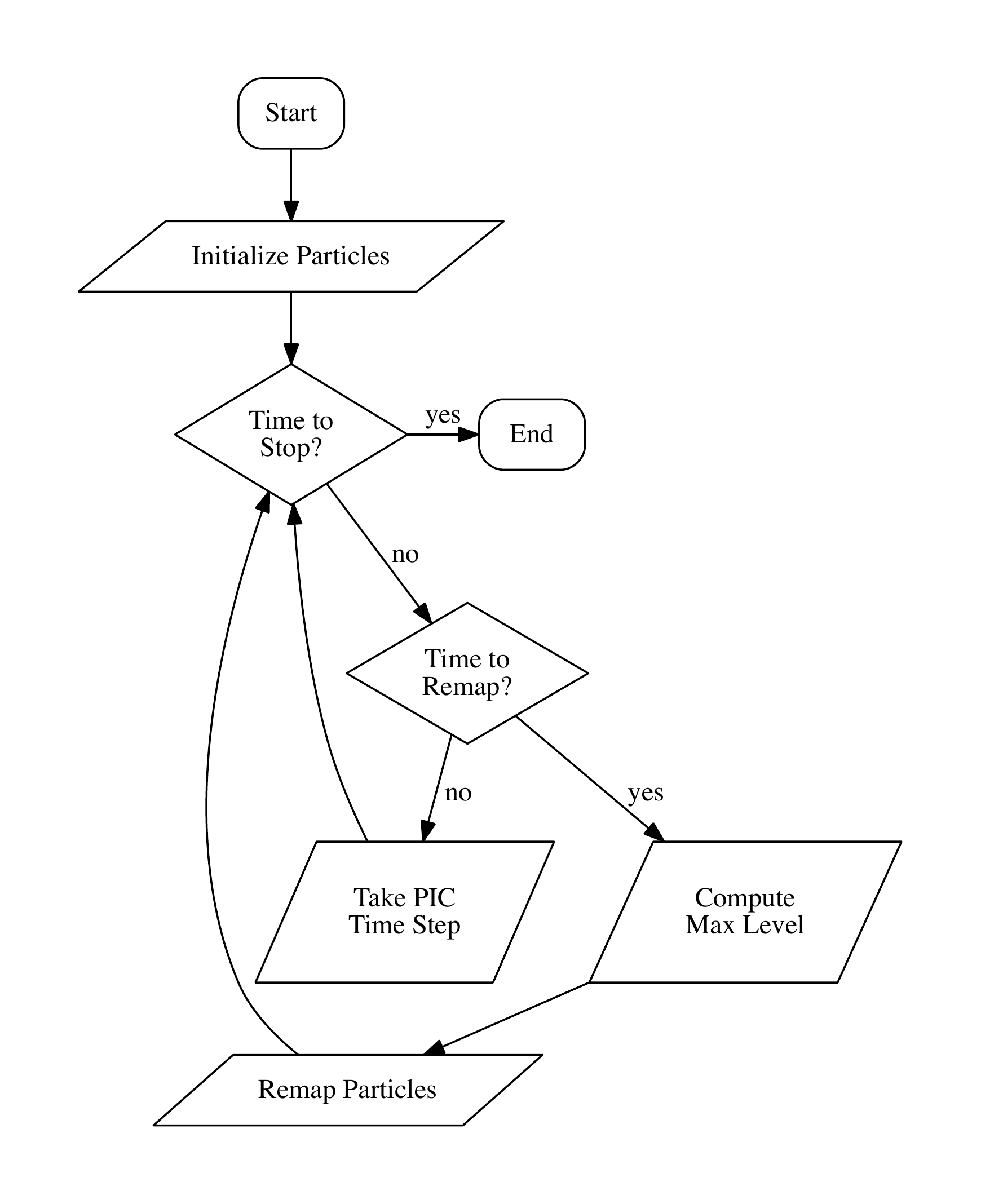}
\caption{The overall control flow of our PIC algorithm, with remapping.}
\label{control}
\end{figure}

\subsection{Remapping on a Uniform Mesh}

As an illustration, consider remapping a set of particles on $\Omega_0$ only. With only one level, our remapping algorithm is identical to \citetalias{wang_particle--cell_2011}. The remapping step proceeds as follows:

\textbf{Deposition.} From the particle data, we represent $f(\bi{x}, \bi{v}$) on $\Omega_0$ by deposition. The position and velocity coordinates of cell centers on $\Omega_0$ are the same as those used for initialization, $\bi{x}_i = (\bi{i} + 1/2) h_x$ and $\bi{v}_j = (\bi{j} + 1/2) h_v - V$, with ($\bi{i}, \bi{j}) \in (\mathbb{Z}^D, \mathbb{Z}^D)$. The values of $f$ on the mesh points are thus:
\begin{equation}
f_{\bi{i} \bi{j}}  = 
\sum_p \left( \frac{m_p}{\Gamma} \right) \bi{W}_4 \left( \frac{\bi{x}_\bi{i} - \bi{x}_p}{h_x} \right) \bi{W}_4 \left( \frac{\bi{v}_{\bi{j}} - \bi{v}_p}{h_v} \right),
\end{equation} where $\Gamma = h_x^D h_v^D$ is the phase-space cell volume and $\bi{W}_4(\bi{x})$ is the $2 D$-dimensional version of the third-order accurate interpolating function from \cite{monaghan_particle_1985}:
\begin{equation}
\bi{W}_4 \left(\bi{x} \right) = \prod_{d = 1}^{D} W_{4} \left( x_d \right),
\end{equation}
\begin{equation}
   W_4(x) =
   \left\{
     \begin{array}{lr}
       1 - \frac{5}{2} \lvert x \rvert^2 + \frac{3}{2} \lvert x \rvert^3, & 0 \leq \lvert x \rvert \leq 1, \\
       \frac{1}{2}\left( 2 - \lvert x \rvert^2 \right) \left( 1 - \lvert x \rvert \right), & 1 \leq \lvert x \rvert \leq 2, \\ 
       0 & \text{otherwise.}
     \end{array}
   \right.
\end{equation} High-order interpolation is necessary because one order of accuracy is lost during the remap step \citepalias{wang_particle--cell_2011}. Thus, for the overall scheme to have second-order accuracy, a third-order or higher interpolating function must be used.

\textbf{Positivity Preservation.} A consequence of high-order interpolation is that $f$ is not guaranteed to be positive for all cells in $\Omega_0$. To account for this, a correction, $\Delta f$, must be applied to the distribution function after it is deposited onto the mesh. We accomplish this using a mass transfer procedure that redistributes matter to cells with $f < 0$ from neighboring cells in proportion to each neighbor's current value of $f$. We write the undershoot in cell ($\bi{i}, \bi{j}$) as $\delta f_{\bi{i} \bi{j}}$:
\begin{equation}
\delta f_{\bi{i} \bi{j}} = \min(0, f_{\bi{i} \bi{j}}),
\end{equation} and initialize $\Delta f = 0$. For every cell with negative $\delta f$, we compute the capacity of each neighboring cell $\mathcal{C}_{\bi{i+m}, \bi{j+n}}$ as
\begin{equation}
\mathcal{C}_{\bi{i+m}, \bi{j+n}} = \max(0, f_{\bi{i+m}, \bi{j+n}}),
\end{equation} where the notation ($\bi{i} + \bi{m}, \bi{j} + \bi{n})$ refers to any neighboring cell in the redistribution zone. We perform this redistribution over a window of $\mathcal{N}$ cells in each direction. The value of $\mathcal{N}$ must be chosen so that it matches the number of cells involved in the particle interpolation stencil: in this case, $\mathcal{N} = 4$. The total capacity is 
\begin{equation}
\mathcal{C}_{\text{tot}} = \sum_{m' \ne 0, n' \ne 0}^{\text{neighbors}} \mathcal{C}_{\bi{i+m'}, \bi{j+n'}}.
\end{equation}

The correction to $f$ within the redistribution window is incremented as
\begin{equation}
 \Delta f_{\bi{i+m}, \bi{j+n}} \mathrel{+}= \frac{\mathcal{C}_{\bi{i+m}, \bi{j+n}}}{\mathcal{C}_{\text{tot}}} \delta f_{\bi{i} \bi{j}}.
\end{equation} This procedure is repeated for each non-positive cell in $\Omega_0$, and the correction is then applied to $f$. Note that there is no guarantee that $f$ will be positive after just one pass of this algorithm. In general, we must iterate until $f$ is positive everywhere. The full positivity preservation algorithm looks like:

\begin{algorithmic}
\WHILE{max($f$) $>$ $0$} 
\STATE{
Set $\Delta f = 0$
\FORALL{$(\bi{i}, \bi{j}) \in \Omega_0$} 
\STATE{
\IF{$f_{\bi{i} \bi{j}}$ $<$ $0$} 
\STATE{

\FORALL{neighbors $(\bi{i + m}, \bi{j + n})$} 
\STATE{
\begin{equation}
 \Delta f_{\bi{i+m}, \bi{j+n}} \mathrel{+}= \frac{\mathcal{C}_{\bi{i+m}, \bi{j+n}}}{\mathcal{C}_{\text{tot}}} \delta f_{\bi{i} \bi{j}}.
 \end{equation}
} \ENDFOR
} 
\ENDIF
} 
\ENDFOR
} 
\begin{equation}
 f^{\text{new}} =  f^{\text{old}} + \Delta f.
\end{equation}
\ENDWHILE
\end{algorithmic} In practice, we find that 2 or 3 iterations is usually sufficient to ensure that the positivity of the distribution function is preserved. 

\textbf{Particle generation.} Finally, we generate a new set of particles by sampling the remapped distribution at every cell $(\bi{i}, \bi{j})$ in $\Omega_0$ and initializing a particle with mass 
\begin{equation}
m_p = f_{\bi{i} \bi{j}} h_x^D h_v^D.
\end{equation}  As in the problem initialization, we discard particles with masses less than $10^{-12}$. 

\subsection{Remapping on a Fixed Hierarchy}

\begin{figure*}
\plotone{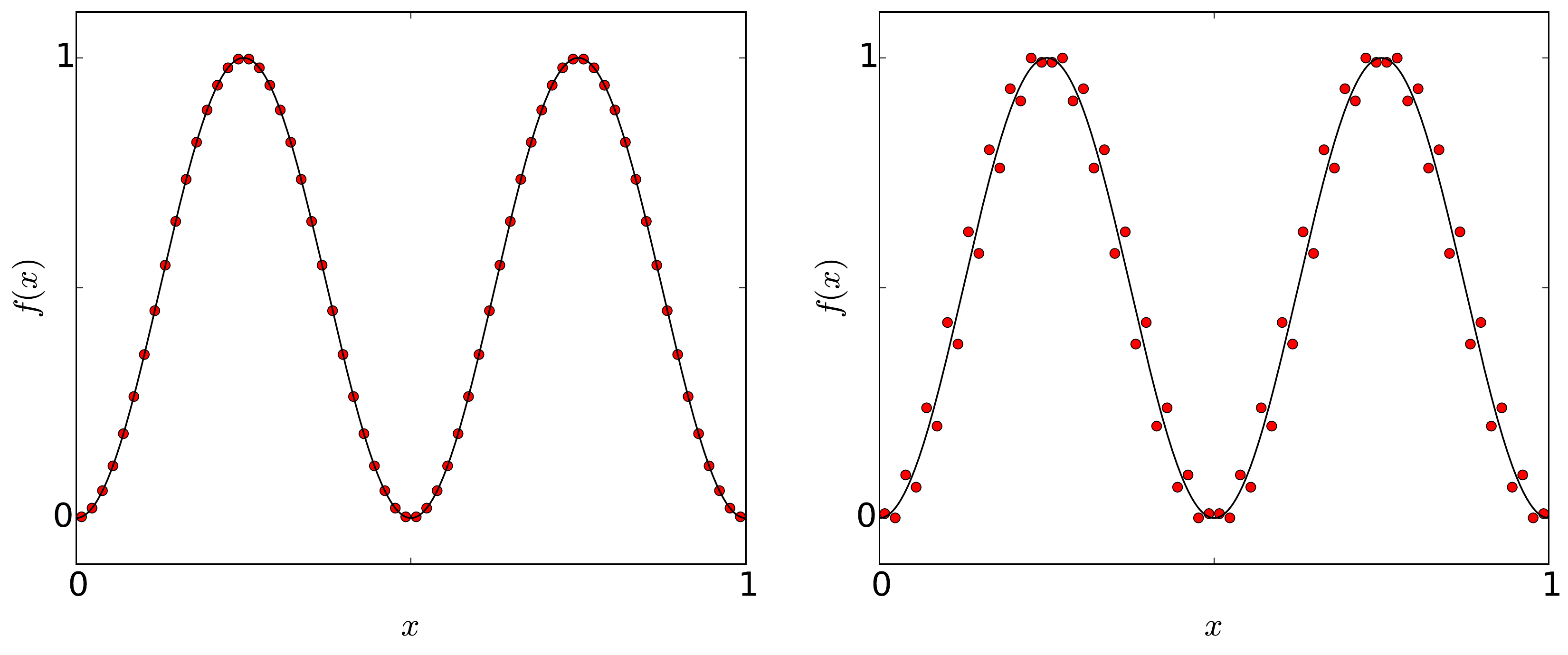}
\caption{A 1D demonstration of our method for depositing particle quantities on refined meshes. In both panels, an initial sinusoidal function $f(x)$ (the black curve) has been sampled by a set of 32 particles, such that $h_x = 1/32$. These particles were then deposited onto a 64 cell mesh ($\Delta x = 1/64$). This mesh was used to generate a new set of particles. In the left panel, the coarser spacing $h_x$ was used for deposition, while in the right panel, the fine cell spacing $\Delta x$ was used instead. The coarse spacing results in a much smoother representation of the underlying function.}
\label{interpolation}
\end{figure*}

Next, consider the case with a hierarchy of levels, \{$\Omega_0$, $\Omega_1$, ..., $\Omega_{\ell_{\text{max}} - 1}$\}. We impose the constraint that these levels be properly nested by a buffer of $n_{\text{buff}} = 4$ fine-level cells; this choice is determined by the above interpolation stencil. For now, we assume that the levels are fixed in time and known in advance; we consider the full AMR case in the next section. We must now compute $f$ on the composite grid. To do so, we first partition the particles $\mathbb{P}$ into sets \{$\mathbb{P}_0$, $\mathbb{P}_1$, ..., $\mathbb{P}_{\ell{\text{max}} - 1}$\} according to what level they will be deposited on.  We perform this partition by associating each particle with the finest level that contains its entire interpolation stencil; particles that lie too close to a coarse/fine boundary for their entire clouds to be contained on the fine level go on the next coarser level.

Remapping on a hierarchy of meshes raises the additional issue of what mesh spacing to use in the deposition step: the spacing of the mesh that the particles are being interpolated \emph{to}, or the spacing of the mesh the particles were generated $\emph{from}$. It is common for AMR PIC schemes to use the spacing of the target mesh for density deposition. However, we find that this technique leads to spurious oscillatory features in the distribution function when used in our remapping algorithm (see Figure \ref{interpolation}). Instead, we find it necessary for each particle to remember the spacing of the mesh on which it was generated. We label these spacings $\epsilon_x^p$ and $\epsilon_v^p$. 

With that caveat, the remapping step proceeds as follows:

\textbf{Deposition.} We deposit the particles in each set onto the appropriate level. For example, on $\Omega_0$:
\begin{align}
f_{\bi{i} \bi{j}}^0 &= \nonumber \\
& \sum_{p \in \mathbb{P}_0} \left( \frac{m_p}{\Gamma} \right) \bi{W}_4 \left( \frac{\bi{x}_\bi{i} - \bi{x}_p}{ \delta x^0} \right) \bi{W}_4 \left( \frac{\bi{v}_{\bi{j}} - \bi{v}_p}{\delta v^0} \right),  \hspace{0.05 in} \nonumber \\
& \text{for} \hspace{0.05 in} (\bi{i}, \bi{j}) \in \Omega_0, 
\end{align} where the expression is evaluated for \emph{every} cell $(\bi{i}, \bi{j})$ in $\Omega_0$, not just the valid cells. The quantities $\delta x^\ell$ and $\delta v^\ell$ are:
\begin{align}
\delta x^\ell &= \max(h_x r_{\text{ref}}^\ell, \epsilon_x^p ) \nonumber \\
\delta v^\ell &= \max(h_v r_{\text{ref}}^\ell, \epsilon_v^p ).
\end{align} That is, we use whichever is larger, the mesh spacing of the target mesh or the mesh spacing with which the particles were generated. We find that this interpolation scheme results in a smooth representation of the underling distribution function even when adaptive refinement is employed.

The corresponding expression for level 1 is:
\begin{align}
f_{\bi{i} \bi{j}}^1 &= \nonumber \\
& \sum_{p \in \mathbb{P}_1} \left( \frac{m_p}{\Gamma} \right) \bi{W}_4 \left( \frac{\bi{x}_\bi{i} - \bi{x}_p}{\delta x^1} \right) \bi{W}_4 \left( \frac{\bi{v}_{\bi{j}} - \bi{v}_p}{\delta v^1} \right) + f_{\bi{i}^c \bi{j}^c}^0, \hspace{0.05 in} \nonumber \\
& \text{for} \hspace{0.05 in} (\bi{i}, \bi{j}) \in \Omega_1. 
\end{align}

The second term in this expression accounts for the particles in $\mathbb{P}_0$ that deposited some of their mass on the region covered by $\Omega_1$. The indices $\bi{i}^c$, $\bi{j}^c$ are \emph{coarsened} versions of ($\bi{i}$, $\bi{j}$):
\begin{align}
\bi{i}^c &= \left( \left\lfloor \frac{i_0}{r_{\text{ref}}}\right\rfloor, ..., \left\lfloor \frac{i_{D-1}}{r_{\text{ref}}}\right\rfloor \right), \nonumber \\ 
\bi{j}^c &= \left( \left\lfloor \frac{j_0}{r_{\text{ref}}}\right\rfloor, ..., \left\lfloor \frac{j_{D-1}}{r_{\text{ref}}}\right\rfloor \right).
\end{align}

Note that by adopting a 4-cell proper nesting requirement, we avoid the case where this coarse/fine correction can involve more than 2 levels. This process continues up to $\ell_{\text{max}}$. In general:
\begin{align}
f_{\bi{i} \bi{j}}^\ell &=  \nonumber \\
& \sum_{p \in \mathbb{P}_\ell} \left( \frac{m_p}{\Gamma} \right) \bi{W}_4 \left( \frac{\bi{x}_\bi{i} - \bi{x}_p}{\delta x^\ell} \right) \bi{W}_4 \left( \frac{\bi{v}_{\bi{j}} - \bi{v}_p}{\delta v^\ell} \right) + f_{\bi{i}^c \bi{j}^c}^{\ell - 1}, \nonumber \\
& \hspace{0.05 in} \text{for} \hspace{0.05 in} (\bi{i}, \bi{j}) \in \Omega_\ell. 
\end{align} The result of applying this procedure over levels \{$\Omega_0$, ..., $\Omega_{\ell_{\text{max}} - 1}$\} is a representation of $f$ that fully conserves matter over the composite grid. 

\textbf{Positivity Preservation.} The positivity preservation step is similar to the above. The difference is that when applying positivity preservation algorithm to level $\ell$, we must sometimes draw mass from cells that lie on level $\ell + 1$ and $\ell - 1$. For convenience, we accomplish by augmenting each level with a later of of $\mathbb{N}$ ghost cells. When we apply the correction to $f^{\ell}$, we also apply it to the subsets of levels $\ell + 1$ and $\ell - 1$ that are covered by the level $\ell$ ghosts. To coarsen the correction, we use simple arithmetic averaging, and to refine it, we use piecewise-constant interpolation.   

\textbf{Particle Generation.} The particle generation step proceeds as in the single-level case, except that we generate one particle for every \emph{valid} cell in the hierarchy. That is, we do:
\begin{algorithmic}
    \FOR{$\ell = 0$, ..., $\ell_{\text{top}} - 1$}
    \STATE{
    $ $
    \FORALL{$(\bi{i}, \bi{j}) \in \Omega_{\ell, \text{valid}}$} 
    \STATE{
    Initialize particle with mass $m_p = f_{\bi{i} \bi{j}}^\ell h_x^D h_v^D.$
    }
    \ENDFOR
    }
    \ENDFOR
\end{algorithmic}
    
    \subsection{Remapping with AMR}
    
\begin{figure*}
\epsscale{1.2}
\plotone{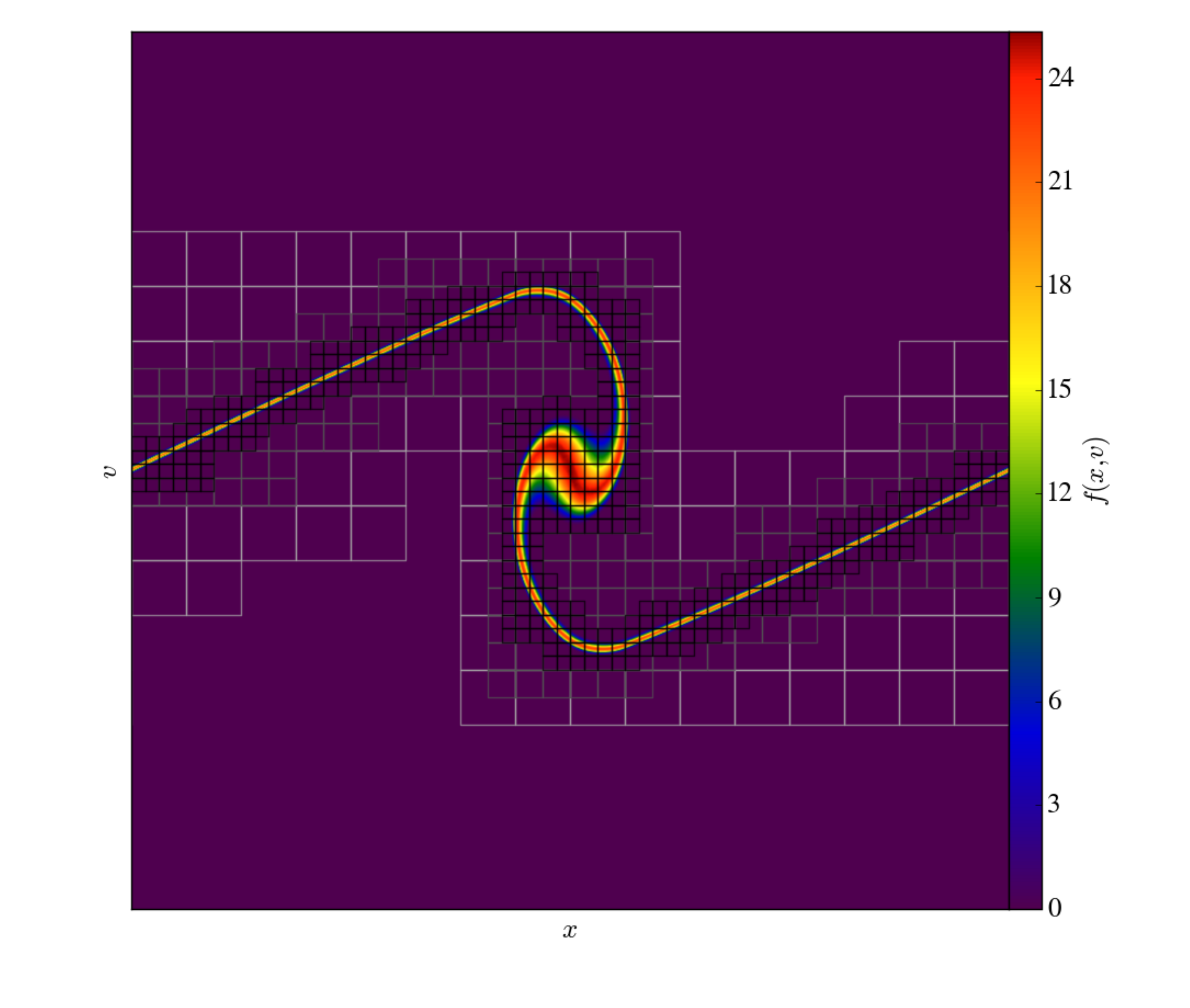}
\caption{An example of a 1 + 1D distribution function represented on an adaptive hierarchy of grids. The level 0 grids have been omitted for clarity, while the level 1, 2, and 3 grids have been over-plotted in white, gray, and black, respectively. }
\label{f_with_grids}
\end{figure*}
    
    In practice, the AMR hierarchy is not known at the beginning of a remapping step. Instead, the AMR level structure must be self-consistently built up from the particle distribution and a suitable set of refinement criteria. In this paper, we use the condition that a cell is tagged for refinement whenever the value of $f_{\bi{i} \bi{j}}^\ell$ exceeds the predetermined threshold $f_{\text{thresh}} = 0.1.$ Finer levels are then generated in such a way that they cover these tags, and are consistent with our proper nesting requirement. The other input to the process is the maximum number of allowed levels, $\ell_{\text{max}}$. We then build up the AMR representation of $f$ using the standard bootstrapping procedure:
    
    \begin{algorithmic}
    \FOR{$\ell_{\text{top}} = 0$, ..., $\ell_{\text{max}} - 1$} 
    \STATE{ 
    Partition particles into \{$\mathbb{P}_0$, ..., $\mathbb{P}_{\ell_{\text{top}}}$\} \\
    Deposit particles on \{$\Omega_0, ..., \Omega_{\ell_{\text{top}}}$\} \\
    \IF{$\ell_{\text{top}} < \ell_{\text{max}} - 1$} 
    \STATE{
    Regrid:
    \FOR{$\ell = 0$, ..., $\ell_{\text{top}}$}
    \STATE{
    Tag cells on $\Omega_{\ell}$ where $f_\ell > f_{\text{thresh}}$
    }
    \ENDFOR
    \FOR{$\ell = 1$, ..., $\ell_{\text{top}} + 1$}
    \STATE{
    Generate new $\Omega_{\ell}$
    }
    \ENDFOR
    }
    \ENDIF 
    }
    \ENDFOR
    \end{algorithmic}
    
    The positivity preservation and particle generation steps are the same as for the fixed hierarchy case. The result of this procedure is an AMR representation of $f$. See Figure \ref{f_with_grids} for an illustration in the 1 + 1D case. Note that, since the particle remap is a purely local process, it is not necessary (nor would it be advisable in 3 + 3D) to store the full phase space distribution at once. However, in the present work we confine our attention to low-dimensional problems, so we store the full $f$ for simplicity and for illustrative purposes.
        
    \subsection{Refinement Criteria}
        
    The goal of our use of AMR is to resolve the artificial velocity dispersion $\sigma_i$ by a roughly constant number of particles during the remap step, even as $\sigma_i$ decreases with the inverse of the expansion factor. We achieve this by fixing the velocity-space domain boundaries and the cell spacing of the base grid and adding additional levels of refinement to the remapping mesh as the universe expands. We compute the number of AMR levels by requiring that $\sigma(a)$ is always resolved by at least $N_{\sigma}$ cells. If the cell spacing of the base grid in velocity space is $h_v$, then we perform the remapping on an AMR hierarchy with 
\begin{equation}
\label{refinement_criterion}
\ell_{\text{max}} = \left\lceil \frac{\log \left( N_{\sigma} h_v / \sigma_i \right)}{\log \left( r_{\text{ref}} \right) } \right\rceil
\end{equation} levels of refinement, where $r_{\text{ref}}$ is the AMR refinement ratio. 
                
        \subsection{Summary}
        
        We show the updated control flow of PIC with remapping in Figure \ref{control}. After initialization, we begin PIC time integration using the time step in section \ref{sec:time_stepping}. We continue until we reach either the final expansion factor, $a_{\text{stop}}$, or the expansion factor associated with the first remap. The remapping can be applied either in fixed interval in $a$ or every fixed number of time steps. For each remap step, we compute the finest allowed level $\ell_{\text{max}}$ using Equation \ref{refinement_criterion}, and perform the remap on \{$\Omega_0$, ..., $\Omega_{\ell_{\text{max}}}$ \}. This process continues until $a_{\text{stop}}$ is reached. 
                
\section{The Regularized, Remapped Pancake}
\label{sec:remapped_results}
\subsection{1D Results}
\label{remapped_results_1D}

\begin{deluxetable}{ccccccc}
\tablecaption{\label{1D_table} Summary of parameters for the regularized and remapped 1D pancake runs}
\tablewidth{0pt}
\tablehead{
\colhead{$\sigma_i$} & \colhead{$N_{\text{cells}}$} & \colhead{$N_{\sigma}$} & \colhead{$\Delta a_{\text{remap}}$} &  \colhead{$N_x$} & \colhead{$N_v$} & \colhead{$C_{\text{exp}}$}
}
\startdata
    $1.0$ & $256$ & 2 & 0.01 & $128$ & 128 & $1.0 \times 10^{-2} $ \\
    $1.0$ & $512$ & 4 & 0.01 & $256$ & 256 & $5.0 \times 10^{-3} $  \\
    $1.0$ & $1024$ & 8 & 0.01 & $512$ & 512 & $2.5 \times 10^{-3} $ \\
    $1.0$ & $2048$ & 16 & 0.01 & $1024$ & $1024$ & $1.25 \times 10^{-3} $ \\
\enddata
\end{deluxetable}

\begin{deluxetable}{ccccccc}
\tablecaption{\label{conservation_table} Final energy error as a function of resolution}
\tablewidth{0pt}
\tablehead{
 \colhead{$N_{\text{cells}}$} & Singular \tablenotemark{a}& Regularized \tablenotemark{a} & Remapped \tablenotemark{a}
}
\startdata
    $256$ & $6.3 \times 10^{-4}$ & $3.2 \times 10^{-4}$ & $1.2 \times 10^{-3}$  \\
    $512$ & $1.5 \times 10^{-4}$ & $8.2 \times 10^{-5}$ & $1.5 \times 10^{-4}$  \\
    $1024$ & $4.2 \times 10^{-5}$ & $2.0 \times 10^{-5}$ & $7.1 \times 10^{-6}$  \\
    $2048$ & $1.2 \times 10^{-5}$ & $6.1 \times 10^{-6}$ & $3.3 \times 10^{-6}$  \\
\enddata
\tablenotetext{a}{The displayed quantity is the energy conservation error $\epsilon$, evaluated at $a(t) = 1$. For the regularized and remapped columns, we have used the runs with $\sigma_i = 1$.}
\end{deluxetable}

\begin{figure}
\epsscale{1.2}
\plotone{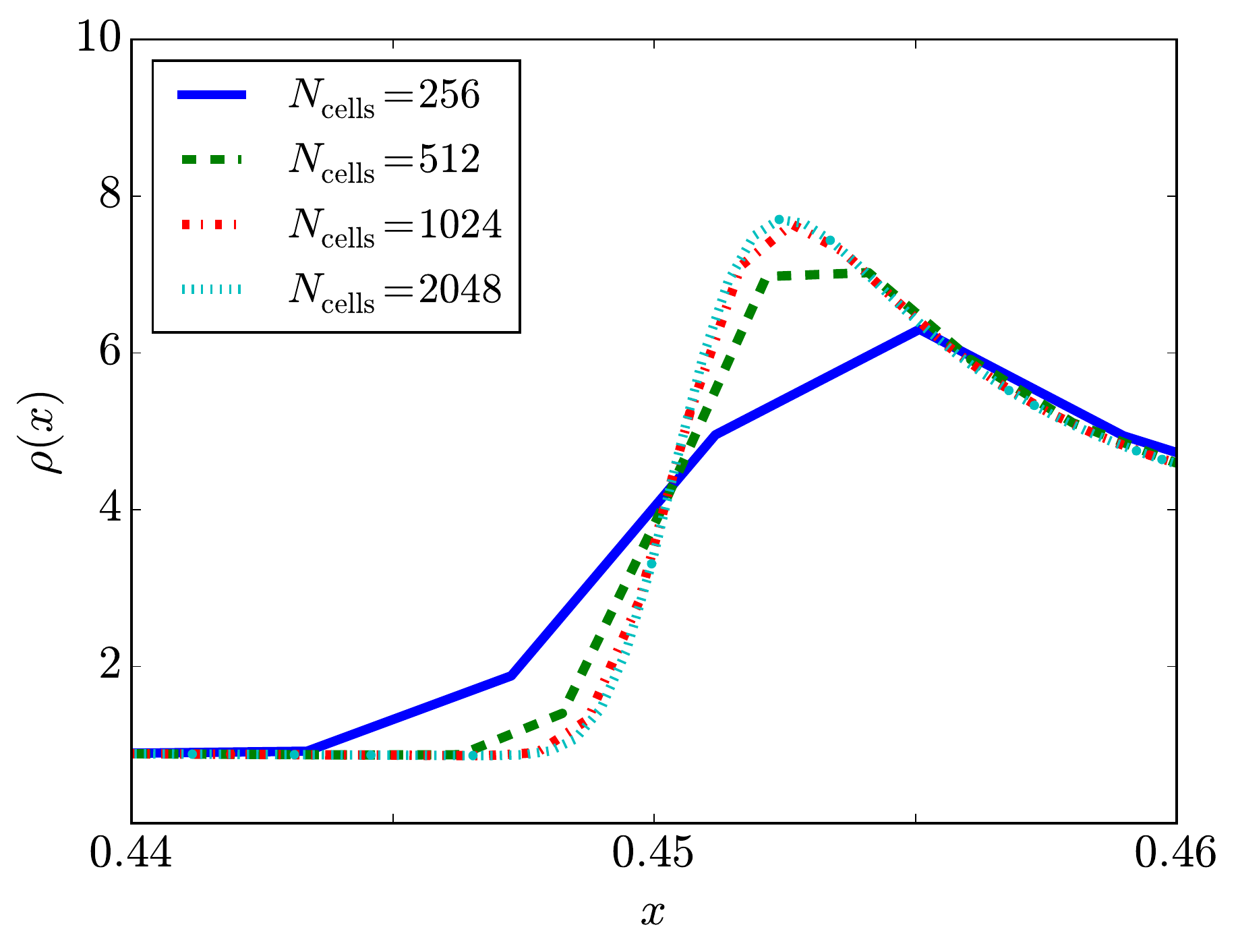}
\caption{A zoomed-in view of one of the outer density caustics from the 1D, regularized and remapped problem with $\sigma_i = 1$, taken at $a(t) = 1$. The different colors correspond to different resolutions: solid blue line - $N_{\text{cells}} = 256$; dashed green line - $N_{\text{cells}} = 512$; dashed-dotted red line - $N_{\text{cells}} = 1024$; dotted cyan line - $N_{\text{cells}} = 2048$. }
\label{1D_density_converge_remapped}
\end{figure}

\begin{figure}
\epsscale{1.2}
\plotone{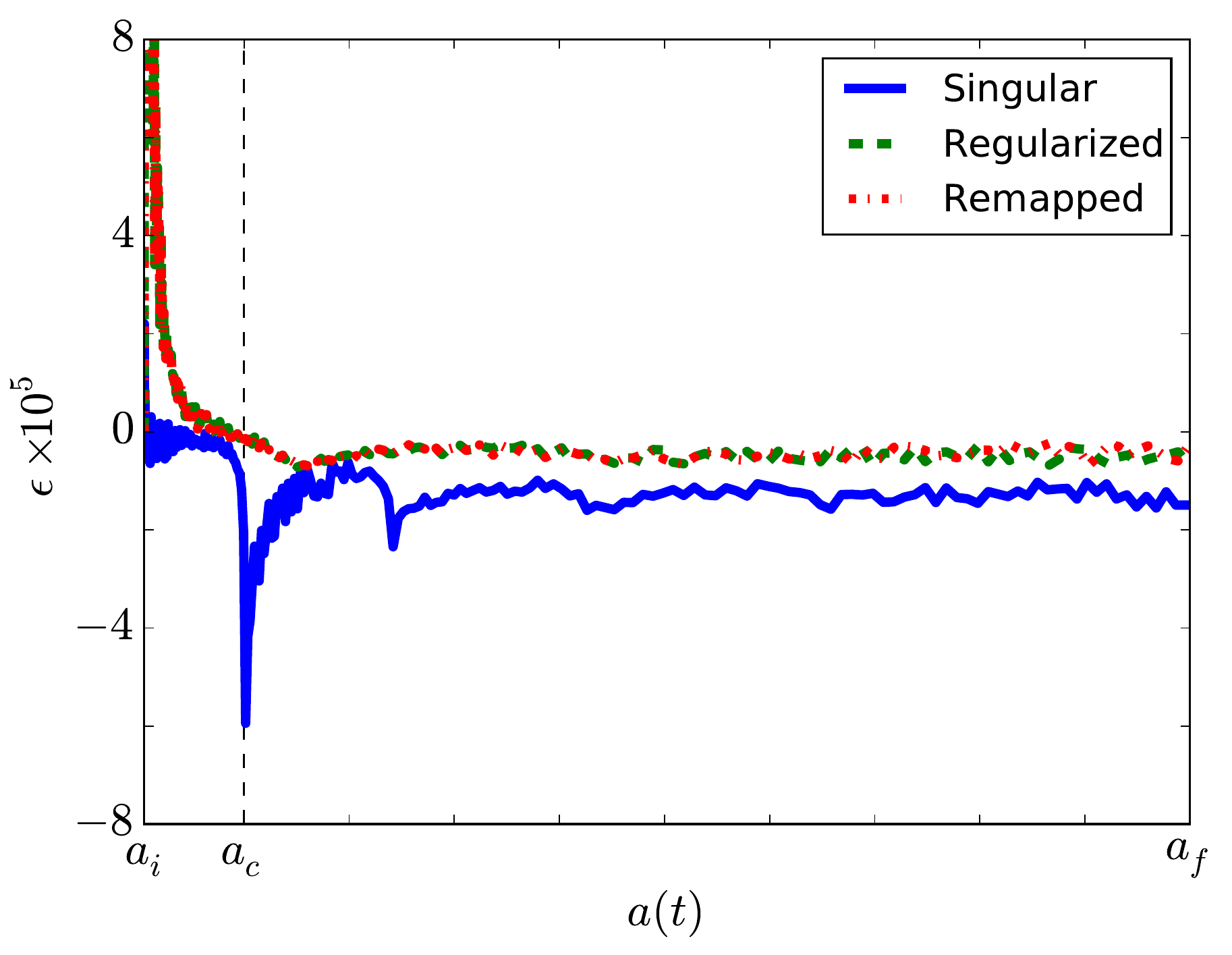}
\caption{The energy conservation error $\epsilon$ (see text for definition) as a function of $a(t)$ for the singular run (blue solid line), the regularized run (green dashed line), and the regularized and remapped run (red dashed-dotted line) for the 1D pancake problem. In this figure, we have used the highest resolution data available, for which $N_{\text{cells}} = 2048$. The variation in the energy error with respect to mesh refinement is summarized in Table \ref{conservation_table}. The vertical dashed line marks the expansion factor at which the first caustic forms.}
\label{energy_conservation}
\end{figure}

\begin{figure*}
\epsscale{1.2}
\plotone{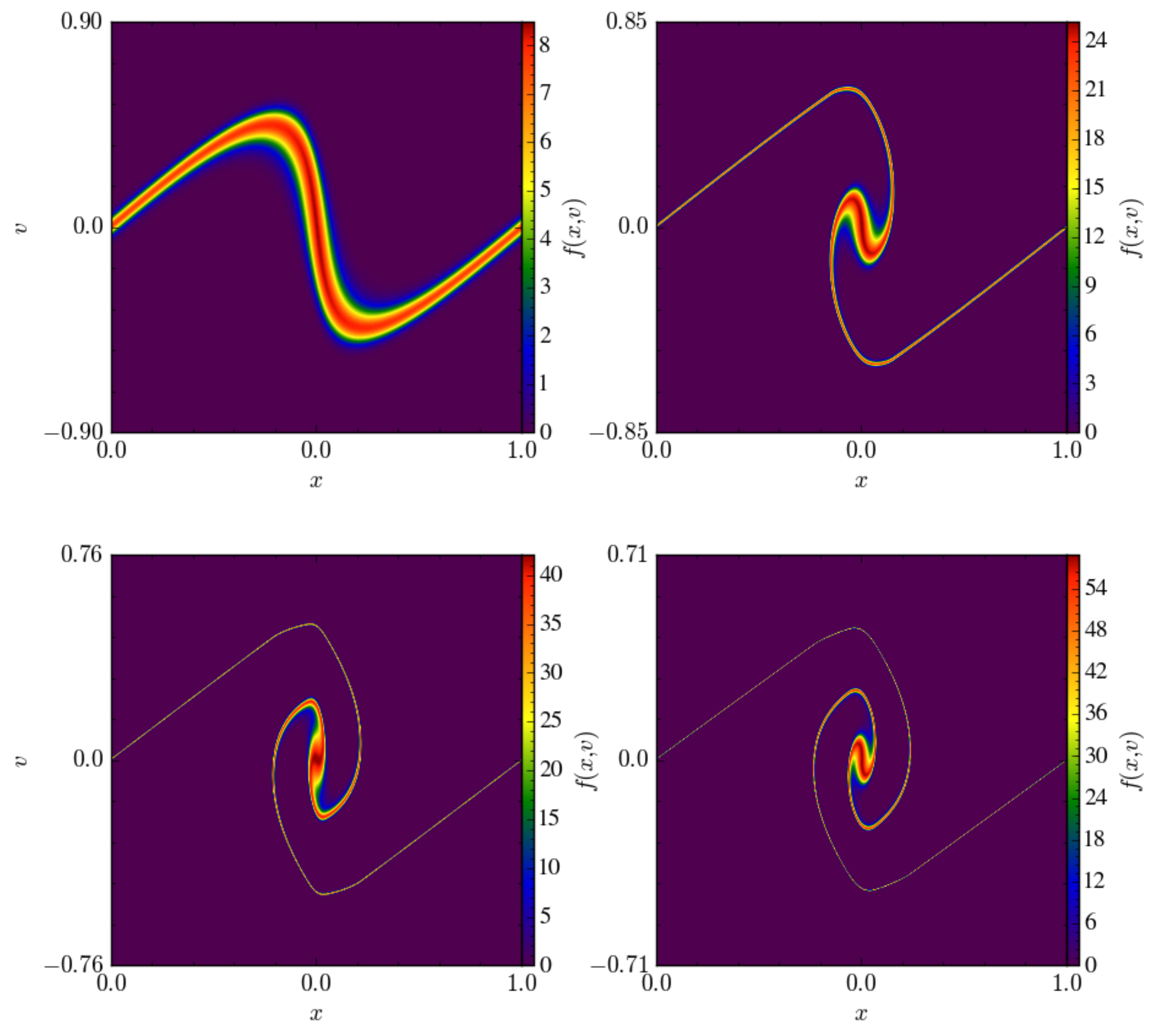}
\caption{Time evolution of the distribution function for the $\sigma_i = 1$ version of the regularized and remapped problem. Top Left - $f(x, v)$ at the expansion factor $a(t) = 0.1$. Top Right - same, at $a(t) = 0.3$. Bottom Left - same, at $a(t) = 0.5$. Bottom Right - same, at $a(t) = 0.7$. The bounds of the images in velocity space have been adjusted slightly in each plot.}
\label{f_with_time}
\end{figure*}

In this section, we examine the performance of the PIC scheme with remapping on the 1D pancake problem with regularized initial conditions (see Section \ref{warm_ics}). Basic PIC converged on this problem (provided the initial conditions were regularized) so we must verify that we can also obtain second order convergence with remapping enabled. For this run, we choose $\sigma_i = 1$, and we arrange our remapping meshes so that this velocity dispersion is always resolved by $N_{\sigma} = 8$ cells. The initial phase-space grid $\Omega_0$ is $N_x \times N_v$, where $N_x = 1 / h_x = 512$, $N_v = 1 / h_v = 512$, and the Poisson mesh is $N_{\text{cells}} = 256$ cells across. By $a(t) = 1$, the mesh spacing in velocity space has been refined by a factor of $2^5$ on the finest level. We apply the remapping every $\Delta a_{\text{remap}} = 0.01$ (i.e. at $a = 0.01,$ $0.02$, etc... ) and dump out $f(x, v)$, $\rho(x)$, $g(x)$, and $\phi(x)$ after every remap. We show the resulting solution at representative times in Figures \ref{1D_density_converge_remapped} through \ref{f_with_time}. In addition to the density, gravitational field, and potential, we have also displayed the full 1 + 1D phase-space distribution function at the same expansion factors. As expected, most of the phase space cells are empty as thus do not need to be represented by a particle.  Visually, the regularized and remapped solutions appear quite similar to the regularization-only solutions, with the effect of $\sigma_i$ being to smooth out the density at the caustic locations so that rigorous convergence becomes possible.

\begin{figure*}
\epsscale{1.2}
\plotone{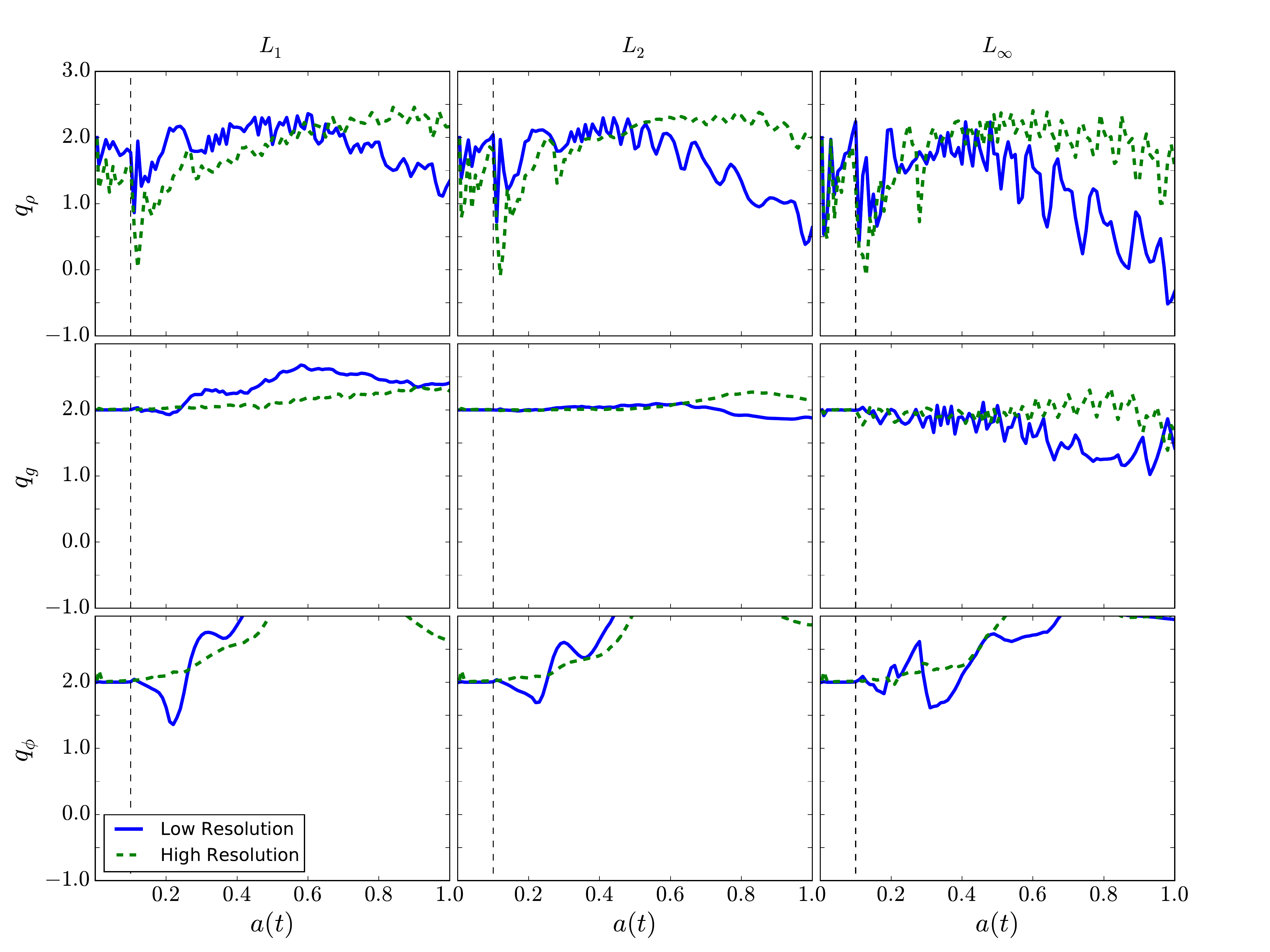}
\caption{The same as Figure \ref{singular_convergence}, but for the $\sigma_i = 1$ version of the regularized, remapped problem. The blue line compares the result for the $N_{\sigma} = 2, 4,$ and $8$ runs, while the green line is for $N_{\sigma} = 4, 8,$ and $16$.}
\label{remapped_convergence}
\end{figure*}

To investigate the convergence rate, we run an analogous set of problems where we vary all of $\Delta_x$, $h_x$, $h_v$, and $\Delta t$ over the same values as in Section \ref{sec:warm_results_1D}. Additionally, we also increase $N_{\sigma}$ by a factor of two as the resolution increases. The resulting convergence rates are plotted in Figure \ref{remapped_convergence}. The convergence rates at late times are much improved compared to the singular runs. Additionally, the convergence rates for the density are improved beyond the regularization-only case. The noisy convergence rates seen in Section \ref{sec:warm_results_1D} at early times for the density are gone. The reason for this behavior is that the remapping procedure constrains the density field to be sampled by several particles per Poisson cell everywhere, which results in a smoother representation of the density. Additionally, the convergence rates for the density are better at late times, as well, with rates close to 2 being obtained in all the norms. The faster-than-second-order convergence rates in the potential are a consequence of the positivity preservation procedure - they are not present if this part of the algorithm is disabled.

It is important to note that, while the above remapping scheme conserves the total $f$ exactly (modulo our choice to discard particles with masses less than $10^{-12}$), it is not guaranteed to conserve total energy, due to discretization error in performing the remapping on a mesh with finite $h_x$ and $h_v$. However, if enough resolution is used, these errors can be made negligibly small. To demonstrate this, we evaluate the overall energy conservation error in our runs as a function of time both with and without remapping. We define the error in energy conservation as follows. If our code conserved energy exactly, then the particles would obey the Layzer-Irvine equation \citep[e.g.,][]{peebles_principles_1993}:
\begin{equation}
\frac{d}{dt} \left[ a (T + U) \right] = -\dot{a} T,
\end{equation} where $T = \frac{1}{2} \sum_{p} m_i v_i^2$ is the total kinetic energy of the particles and $U = \frac{1}{2} \sum_{p} m_i \phi_i$ is their total gravitational potential energy. We use this equation to define the energy conservation error as $\epsilon$:
\begin{equation}
\epsilon = \frac{a(T+U) - a_0 (T_0 + U_0) + \int_{a_0}^{a}T da}{a_0 U_0 - a U}.
\end{equation}

This quantity would be zero in the case of perfect energy conservation. In Figure \ref{energy_conservation}, we compare $\epsilon$ as a function of time for the singular pancake run, the $\sigma_i = 1$ regularized run, and the $\sigma_i = 1$ regularized and remapped run. We use the highest resolution data available ($N_{\text{cells}} = 2048$). Because of our use of a variable time step, neither the singular nor the regularized runs conserves energy exactly, despite our use of a sympletic integrator. However, the overall error in the energy conservation is small - at then end of all three runs, we have conserved energy to better than 0.002 \%. Interestingly, the energy error is actually largest at the end of the singular run, which  suffers a spike in the energy error at first caustic formation.

The energy error is practically the same between the regularized and remapped runs - an indication that, at this resolution, the energy error is dominated by something other than discretization error during the remap. For our lower resolution runs, this is not the case. In Table \ref{conservation_table}, we compare the energy error $a(t) = 1$ between the singular run, the regularized run with $\sigma_i = 1$, and the regularized and remapped run with $\sigma_i = 1$ for four different resolutions. We see that on the $N_{\text{cells}} = 256$ run, the energy error is actually the largest ($\approx$ 0.1 \%) in the run with remapping; at this resolution, the discretization error in the remap step is clearly significant. By $N_{\text{cells}} = 512$, $\epsilon$ in the remapped run has become comparable to that in the singular run, and at still higher resolution, it becomes comparable to the error in the regularized run.   

\begin{figure*}
\epsscale{1.2}
\plotone{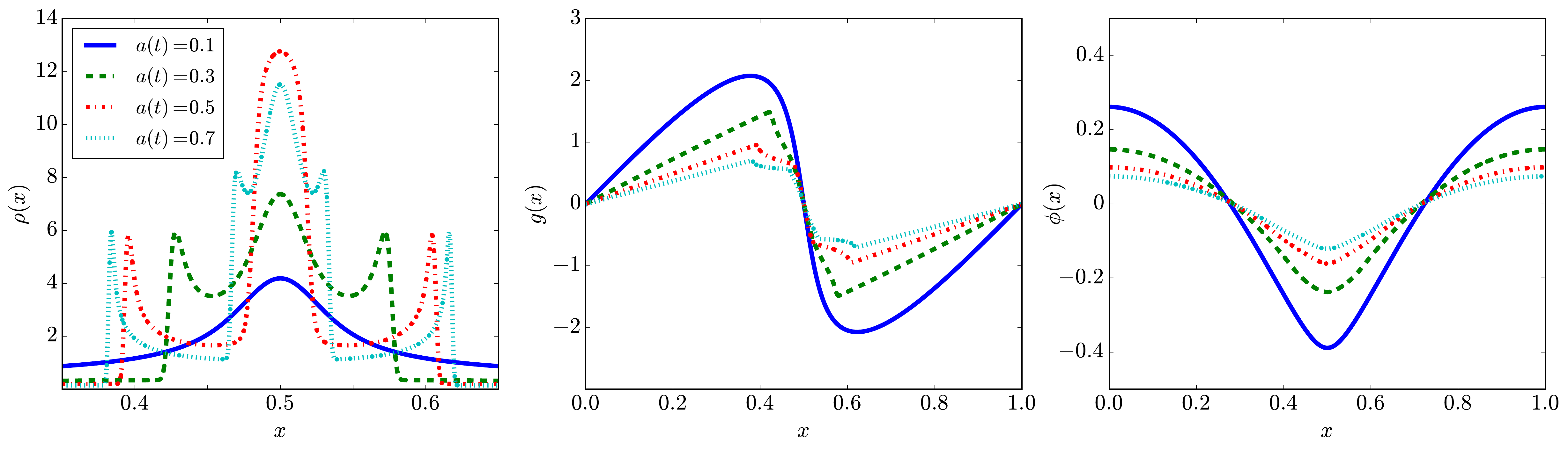}
\caption{The time evolution of the solution to the $\sigma_i = 1$ version of the 1D, regularized and remapped problem. The left panel shows the density, the middle panel the gravitational field, and the right panel the potential. The plotted curves are from the run with $N_{\text{cells}} = 2048$; the $N_{\text{cells}} = 1024$ run would be indistinguishable from the plotted solutions at the scales shown. The different lines correspond to the results at different expansion factors: solid blue line - $a(t) = 0.1$; dashed green line - $a(t) = 0.3$; dashed-dotted red line - $a(t) = 0.5$; dotted cyan line - $a(t) = 0.7$.}
\end{figure*}

\subsection{2D Results}
\label{sec:remapped_results_2D}

\begin{figure*}
\epsscale{1.15}
\plotone{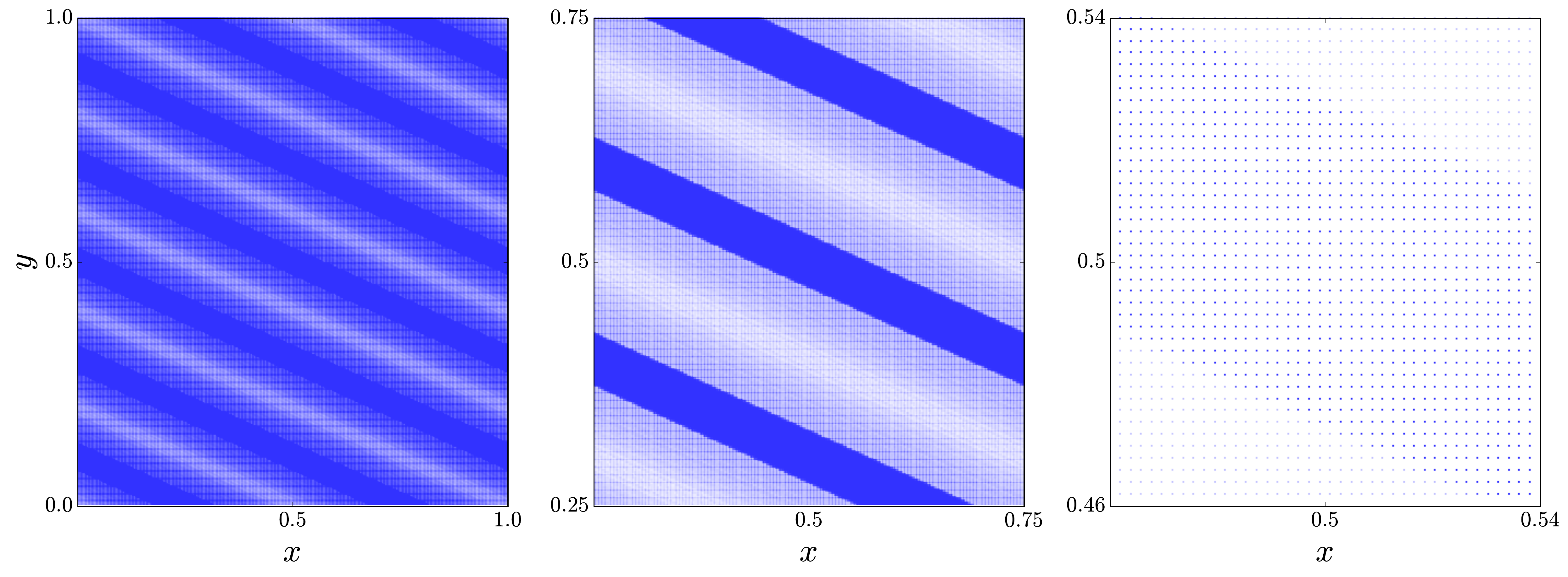}
\caption{Particle $x$ and $y$ positions at $a(t) = 1$ for the $\sigma_i = 1/16$ version of the regularized and remapped oblique pancake problem. The scales shown are the same as in Figure \ref{particle_plot_singular}}
\label{particle_plot_remapped}
\end{figure*}

\begin{figure*}
\epsscale{1.2}
\plottwo{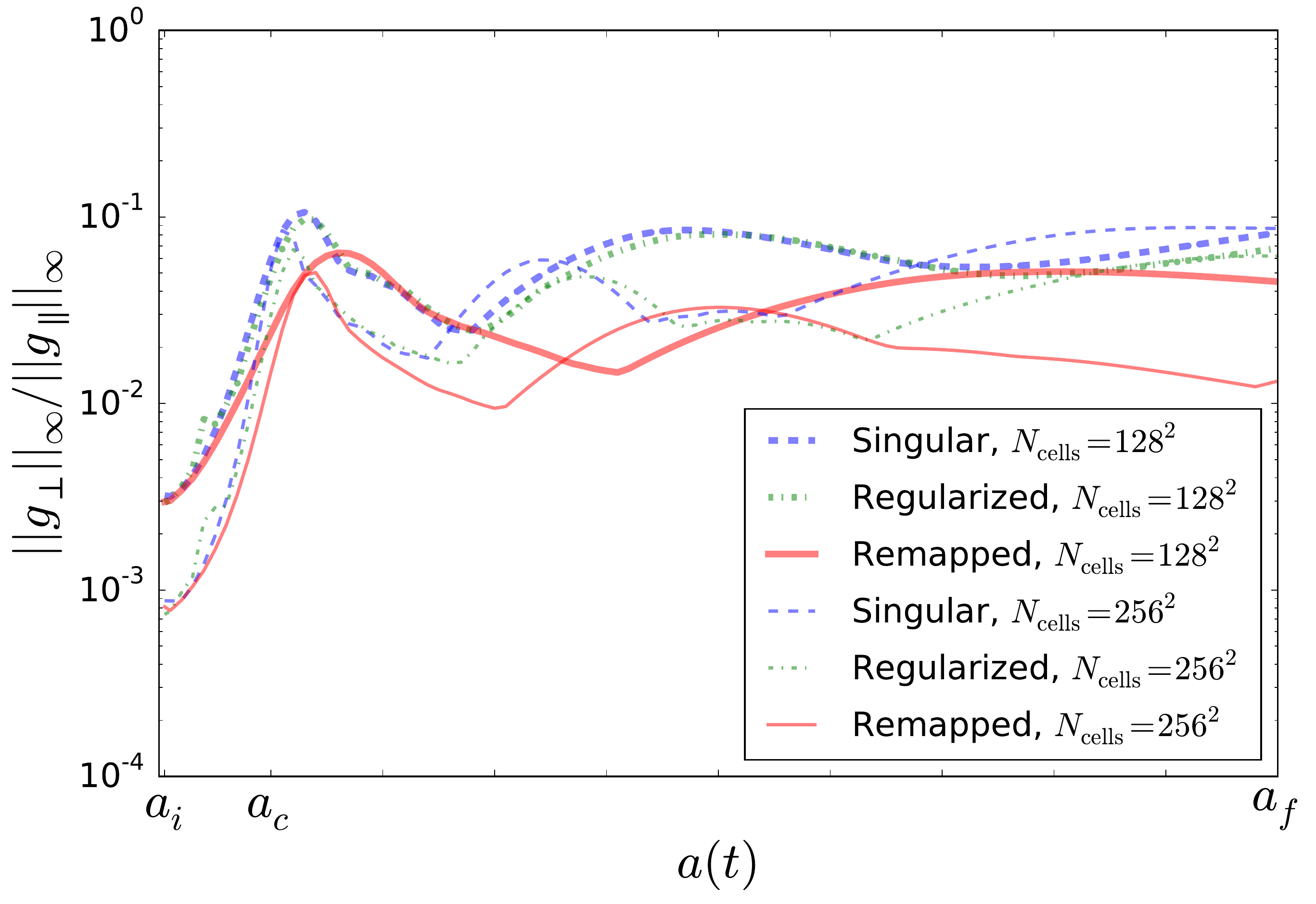}{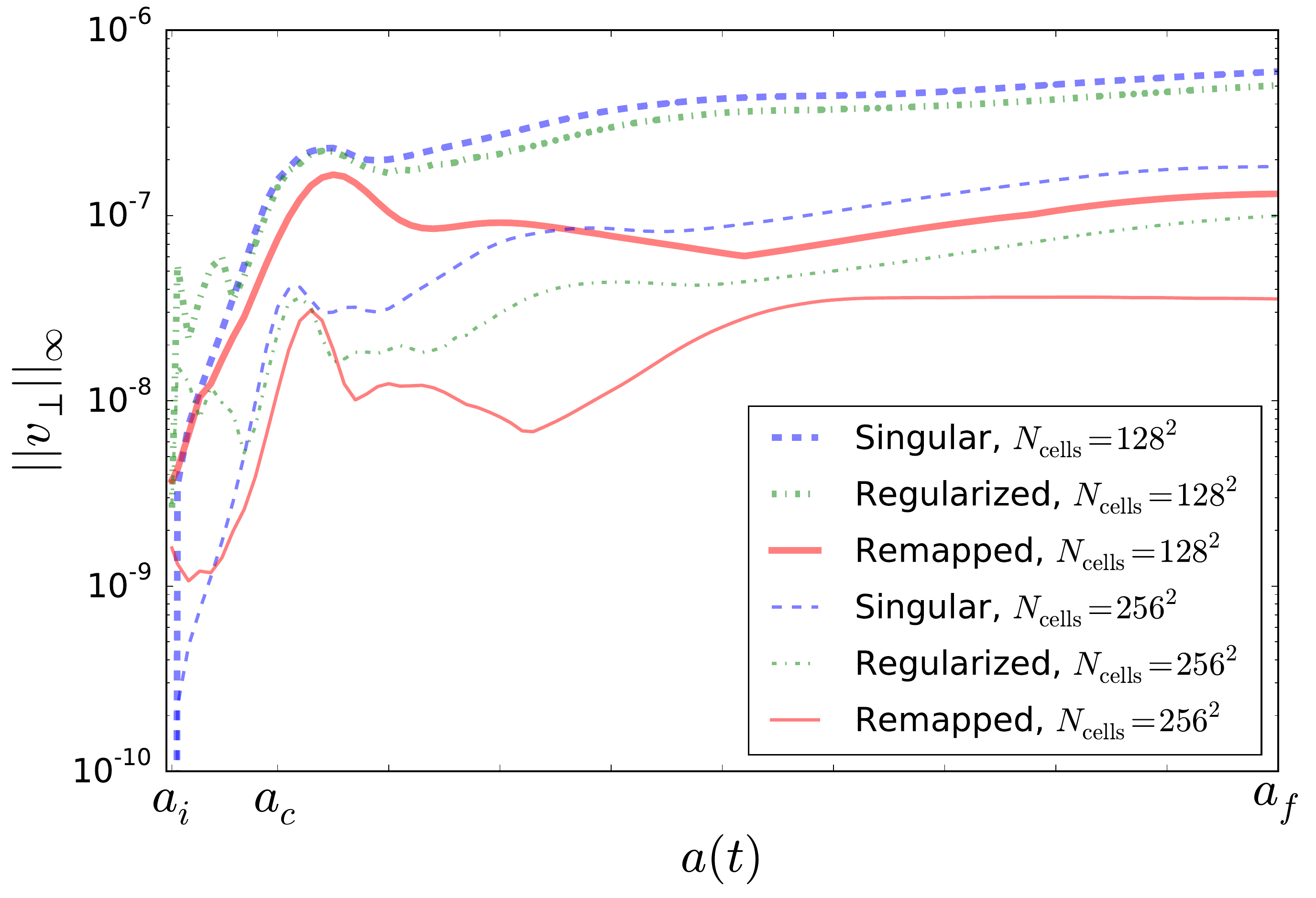}
\caption{Top - A measure of the asymmetry in the gravitational force as a function of time for the singular (dashed blue lines), regularized without remapping (dash-dotted green lines) and regularized with remapping (solid red lines) versions of the oblique pancake problem. Low resolution results are shown using thick lines, and high resolution results are shown using thin lines. The quantity on the y-axis is the maximum off-axis (that is, perpendicular to the perturbation axis) force as computed on the mesh in units of the maximum on-axis (along the perturbation) force. Bottom - a similar plot that shows the growth in the off-axis component of the deposited velocity field; see text for details.}
\label{stability_error}
\end{figure*}

Finally, we repeat the oblique pancake problem with both regularization and remapping turned on. The run parameters are summarized in Table \ref{2D_remapped_table}. We set $\sigma_i = 1/16$ and vary $N_{\text{cells}}$ over \{128, 256\} . We use velocity-space mesh refinement during the remap phase so that $\sigma_i$ is always resolved by at least $N_\sigma = 1$ cells, up until a maximum refinement level of 2 is reached. We remap in linear increments in the expansion factor, with $\Delta a_{\text{remap}} = 0.01$. 

\begin{figure*}
\epsscale{1.2}
\plotone{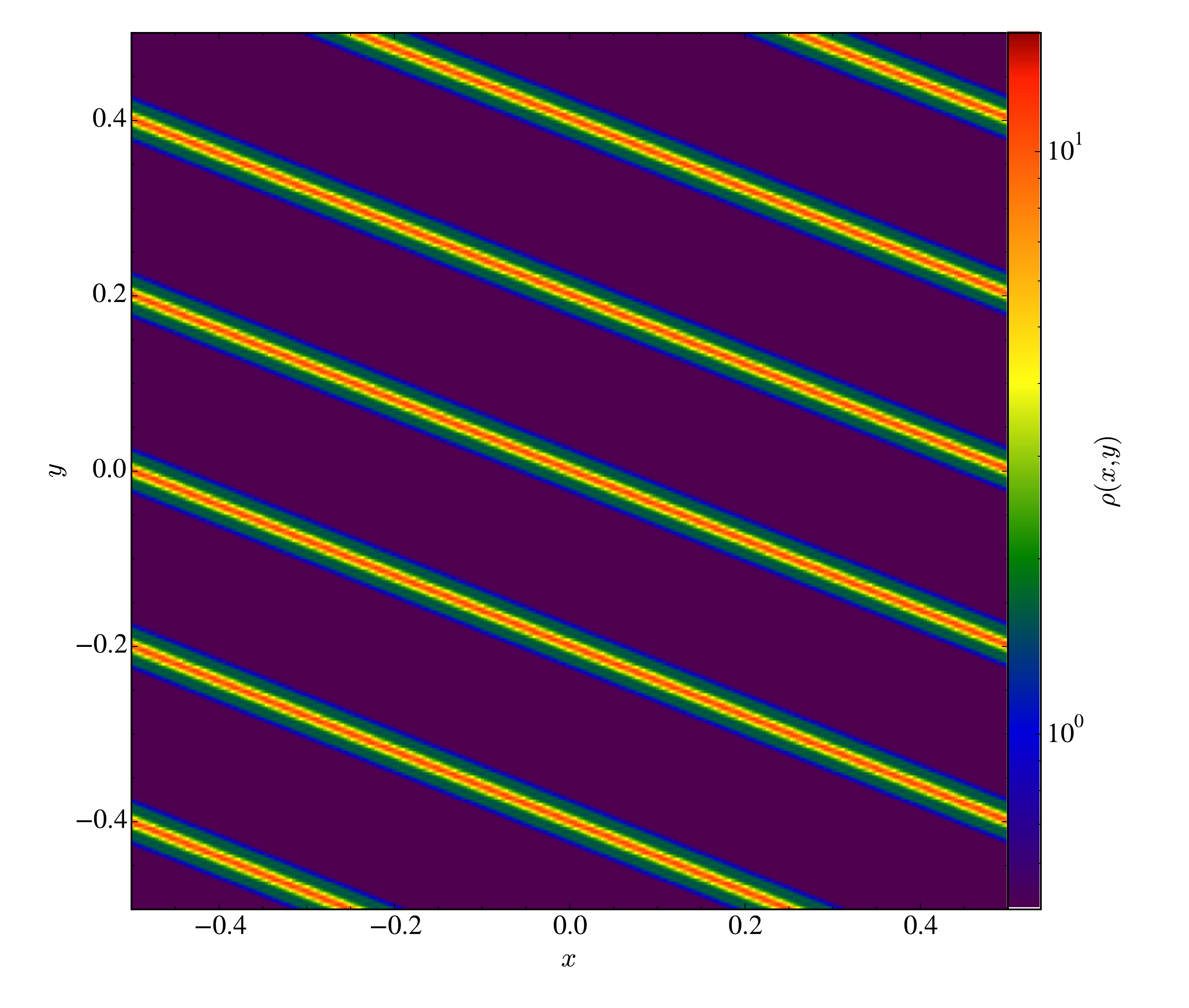}
\caption{The solution for the density at $a=1$ from the low-resolution calculation of the oblique pancake problem using both regularized initial conditions and remapping. The color scale is the same as in Figure \ref{singular}. For details of the parameter choices we adopted, see text.}
\label{remapped}
\end{figure*}

\begin{figure*}
\epsscale{1.2}
\plotone{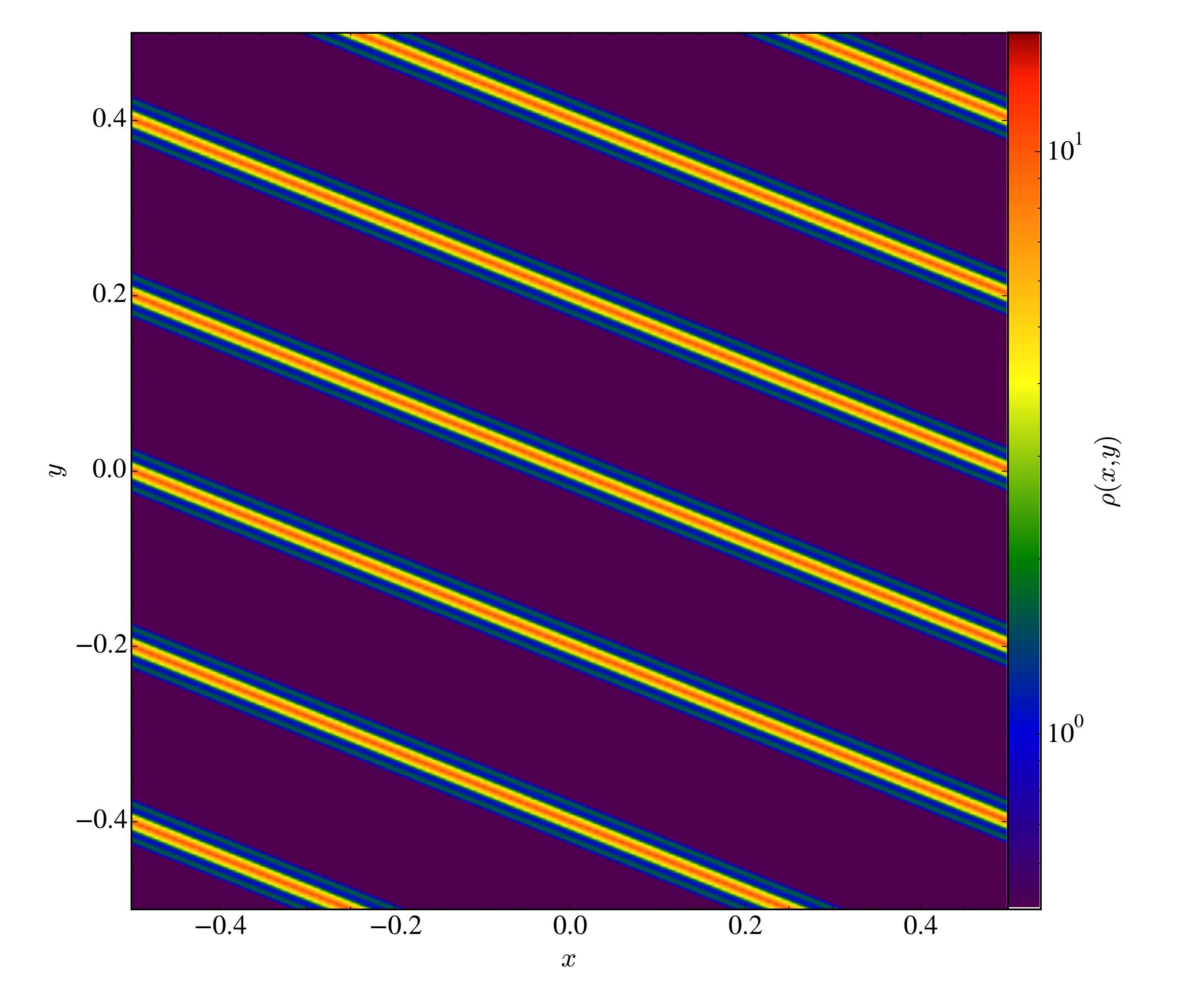}
\caption{The solution for the density at $a=1$ from the high-resolution calculation of the oblique pancake problem using both regularized initial conditions and remapping. The color scale is the same as in Figure \ref{singular}. For details of the parameter choices we adopted, see text.}
\label{remapped_hires}
\end{figure*}

The resulting particle positions are displayed in Figure \ref{particle_plot_remapped}. Unlike Figures \ref{particle_plot_singular} and \ref{particle_plot_regularized}, which show the particles positions at $a=1$ for the singular and regularized but not remapped runs, Figure \ref{particle_plot_remapped} shows no sign of particle clumping along the caustic axes. Instead, a consequence of the remapping procedure is that the particle spacings are forced to be relatively constant throughout the populated regions of phase space. The information about the density is contained in the particle masses, instead. Figures \ref{remapped} and \ref{remapped_hires} shows the grid-deposited density at $a = 1$ for $128^2$ and $256^2$ run, respectively, where we have constructed the 2$D$ density field by depositing the particles onto a $256^2$ and $512^2$ mesh using cloud-in-cell deposition. The unphysical clumping visible in Figures \ref{singular} and \ref{regularized} is no longer visible in these plots.  

It is important to note that this improvement is not simply a matter of using more particles per Poisson cell. Our regularized, remapped calculation of the oblique pancake (Section \ref{sec:remapped_results_2D}) uses approximately 380 particles per Poisson cell on average (the precise number varies with time), almost identical to the number used in the regularized but not remapped run ($\approx 385$), and similar to the 256 used in the corresponding singular calculation. Simply sampling singular initial conditions with more particles in physical space does not lead to a convergent method - our results suggest that the initial conditions must be regularized \emph{and} the extra particles must be arranged in the correct way in phase space.

However, while using many particles per cell is not \emph{sufficient} to prevent artificial fragmentation, our results do suggest that it may be \emph{necessary} in our scheme to carry around more particles per Poisson cell than is currently standard practice. In 3D, even using a relatively modest number of particles in each velocity space direction (say, 4, comparable to our low resolution runs in Section \ref{sec:remapped_results}), would require $4^3 = 64$ particles where one would exist in the comparable singular problem. Clearly, optimizing both the basic PIC scheme and our remapping code for such high particle-per-cell counts will be necessary before our approach can be used for large-scale problems.

We examine our oblique solutions more quantitatively in Figure \ref{stability_error}. Here, we plot the max norm of $g_{\perp}$ over the max norm of $g_{\parallel}$, where $g_{\perp}$ is the component of the gravitational acceleration perpendicular to $\bi{k}$ and $g_{\parallel}$ is the corresponding component parallel to $\bi{k}$. To make a clean comparison between runs, we use the gravitational force as computed on the mesh points by the PIC algorithm in computing this quantity (Equation \ref{PICForce}). In the exact solution, this quantity should be zero, but due to unavoidable discretization errors in the representation of the initial conditions, this will not be precisely true. These errors are largest at the caustic positions, due to the large gradients in the density there. Thus, in all three runs, the degree of asymmetry in the force increases dramatically after $a_{\text{caustic}}$.

To an extent, these errors are an unavoidable consequence of representing a plane-symmetric quantity on a Cartesian grid when that plane is misaligned with the coordinate axes. A stable numerical scheme, however, will limit the ability of these errors to affect the subsequent particle trajectories. As the error analysis of \citetalias{wang_particle--cell_2011} shows, however, PIC is not stable to these errors unless remapping is applied. To investigate this, we have also shown in Figure \ref{stability_error} the growth in the perpendicular component of the deposited velocity, $v_{\perp}$, where the deposited velocity was computed by CIC deposition as in Equation \ref{eq:particle_deposit}, to make a clean comparison between our different runs. As with $g_{\perp}$, we find that $v_{\perp}$ is the largest at the caustic positions. We also find that, in both our low resolution and high resolution models, the application of the remapping procedure reduces these off-axis velocities by almost an order of magnitude. Note that while $v_{\perp}$ is small in absolute terms, both the perpendicular and the parallel components of the deposited velocity should go to zero at the caustic positions, and thus any error in the velocity is relatively large there. Since many particles spend a large amount of time the caustics (because they are turning points of the dark matter distribution function), these small off-axis velocities can lead to significant clumping along the axis of the perturbation, if nothing is done to mitigate them.

\begin{deluxetable}{ccccccc}[b]
\tablecaption{\label{2D_remapped_table} Summary of parameters for the regularized and remapped 2D pancake runs}
\tablewidth{0pt}
\tablehead{
\colhead{$\sigma_i$} & \colhead{$N_{\text{cells}}$} & \colhead{$N_{\sigma}$} & \colhead{$\Delta a_{\text{remap}}$} &  \colhead{$N_x$} & \colhead{$N_v$} & \colhead{$C_{\text{exp}}$}
}
\startdata
    $1/16$ & $128^2$ & 1 & 0.01 & $256$ & 128 & $2.0 \times 10^{-2}$ \\
    $1/16$ & $256^2$ & 1 & 0.01 & $512$ & 128 & $1.0 \times 10^{-2}$ \\
\enddata
\end{deluxetable}

    \section{Conclusions and Caveats}
   \label{sec:conclusions}

In this paper, we have investigated the convergence of PIC schemes on a variety of ``Zel'dovich Pancake" setups. Our conclusions are as follows. In one spatial dimension, the perfectly cold problem does not converge at the formal order of the method at late times, due to singularities in the solution for the density at the positions of the dark matter caustics. This is true regardless of the number of particles per Poisson cell employed. Once these singularities are removed through the introduction of a finite initial artificial velocity dispersion, PIC can converge at the expected order on the 1D problem. Additionally, by taking the limit as $\sigma_i \rightarrow 0$, we have shown that we approach the solution obtained in the perfectly cold limit, except that the peak density are limited by the velocity dispersion. This gives us confidence that, in 1D, the results obtained with singular initial data are in fact valid solutions to the Vlasov-Poisson equation in the zero-temperature limit.
    
On the 2D, oblique pancake problem, however, the basic PIC method with cold initial data does not converge to a valid solution of the Vlasov-Poisson equation. Instead, the solutions exhibit small-scale fragmentation that does not improve with mesh refinement. This problem is not solved by increasing the number of particles per Poisson cell, nor is it solved by using the artificially warm initial data. However, the particle remapping algorithm we describe does improve the accuracy of PIC on this problem, reducing the off-axis component of the gravitational acceleration and velocity significantly when applied 100 times over between $a(t) = 1/200$ and $a(t) = 1$.
    
 Our results show that remapping is primarily important at the caustic positions. This is because that is where the gravitational field varies most rapidly with position - a consequence of the sharply-peaked density structure. Thus, errors in the particle trajectories compound particularly quickly at those locations. Additionally, in more than one spatial dimension, you don't have any control over the directionality of those errors, which leads to the clumping observed in the non-remapped runs. This fact suggests a possible improvement to our remapping scheme: instead of applying it globally to the entire distribution function, we could apply it only at those spatial locations where the density contrast is large.
    
Given that, in 1D, the limiting case of the regularized solutions agrees with the solution obtained using cold initial data, and given that the introduction of $\sigma_i$ does not, by itself, prevent artificial fragmentation in 2D, a possible conclusion is that the artificial velocity dispersion is not necessary in and of itself, but is primarily important in that allows the use of a plasma-style remapping algorithm. If this is the case, it may be possible to construct a remapping algorithm that does not require regularization, by e.g. exploiting the continuity of the distribution function in phase space as in \cite{hahn_new_2013}. However, whether designing such an algorithm is possible is an open question. To use the remapping algorithms that currently exist, regularization is necessary.

There are a number of caveats to our work that bear mentioning. The first is that we have restricted our attention to various configurations of the Zel'dovich Pancake problem. We have not yet applied our technique to other, more realistic test problems or to a full structure formation calculation. Second, we have restricted ourselves to working in one- and two-dimensional spaces only. Work towards creating a three-dimensional version of our remapping algorithm in progress. Finally, while our scheme does use adaptivity in velocity space, we have not yet introduced AMR in the spatial directions, either for solving the Poisson equation or for generating the particle positions during remapping. Such a modification would clearly be advantageous, in that it would concentrate both the particles and the resolution of the force grid towards the regions where high resolution is most needed. However, this was not necessary for the relatively simple problems considered in this paper.  

\acknowledgments{This material is based upon work supported by the U.S. Department of Energy, Office of Science, Advanced Scientific Computing Research Program and performed under the auspices of the U.S. Department of Energy by Lawrence Berkeley National Laboratory under Contract DE-AC02-05CH11231. The code used to run the simulations in this paper was built using Chombo's ParticleTools library, which was originally developed by Francesco Miniati for the \texttt{CHARM} code. AM wishes to thank Anshu Dubey, Daniel Graves, and Daniel Martin for sharing their time and expertise with Chombo, and the anonymous referee for a thoughtful report that improved the quality of the paper. This work relied heavily on yt \footnote{http://yt-project.org} \citep{yt_paper} and the core scientific Python packages, including IPython \footnote{http://ipython.org} \citep{Ipython}, NumPy \footnote{http://www.numpy.org} \citep{NumPy}, and Matplotlib \footnote{http://matplotlib.org} \citep{Matplotlib}, for data analysis and plotting.}    

\bibliographystyle{apj}
\bibliography{cosmologicalpic}
    
\end{document}